\documentclass[a4paper,english,twocolumn,nofootinbib,pre,superscriptaddress,aps]{revtex4-2}
\usepackage{amsmath,amsfonts,bm,physics}
\usepackage{amssymb}
\usepackage{mathrsfs}
\usepackage{graphicx}
\usepackage{xcolor}
\usepackage{hyperref}

\newcommand{\be}{\begin{equation}}
\newcommand{\ee}{\end{equation}}

\newcommand{\ov}{\overline}

\newcommand{\ra}{\rangle}
\newcommand{\la}{\langle}
\newcommand{\bs}{\boldsymbol}
\newcommand{\br}{ {\boldsymbol r} }
\newcommand{\bp}{ {\boldsymbol p} }
\newcommand{\bq}{ {\boldsymbol q} }

\newcommand{\ww}{\widetilde}

\newcommand{\HH}{{\mathcal{H}}}

\begin{document}

\title{
 Semiclassical study of diagonal and offdiagonal 
 functions in the eigenstate thermalization hypothesis
}

\author{Xiao Wang}
\affiliation{ Department of Modern Physics, University of Science and Technology of China,
	Hefei 230026, China}
\affiliation{CAS Key Laboratory of Microscale Magnetic Resonance,
University of Science and Technology of China, Hefei 230026, China}
\author{Wen-ge Wang}\email{wgwang@ustc.edu.cn}
\affiliation{ Department of Modern Physics, University of Science and Technology of China,
	Hefei 230026, China}
\affiliation{CAS Key Laboratory of Microscale Magnetic Resonance,
University of Science and Technology of China, Hefei 230026, China}

\date{\today}

\begin{abstract}

 The so-called eigenstate thermalization hypothesis (ETH), which has been tested in various
 many-body models by numerical simulations, 
 supplies a way of understanding eventual thermalization and
 is believed to be important for understanding processes of thermalization.
 Two functions play important roles in the application of ETH, 
 one for averaged diagonal elements and the other for the variance of 
 offdiagonal elements of an observable addressed by ETH on the energy basis.
 For the former function, a semiclassical expression is known of the zeroth order of $\hbar$,
 while, little is known analytically for the latter. 
 In this paper, a semiclassical expression is derived for the former function,
 which includes higher-order contributions of $\hbar$.
 And, a semiclassical approximation is derived for the latter function,
 under the assumption of negligible correlations among energy eigenfuntions on an action basis.
 Relevance of the analytical predictions are tested numerically in the Lipkin–Meshkov–Glick model.
\end{abstract}

\maketitle

 \section{Introduction}\label{sect-introduction}

 The idea of presuming eigenstate thermalization may be traced back to 
 early works of several authors (see, e.g., Refs.\cite{Deutch91,PRA86-Feing-Peres}).
 While, the first demonstration of this phenomenon in a concrete model
 was given by Srednicki, in a quantum chaotic model --- 
 rarefied hard-sphere gas  \cite{srednicki1994chaos},
 based on the so-called Berry's conjecture given by a semiclassical analysis \cite{Berry77}.
 For a generic many-body quantum chaotic system,
 eigenstate thermalization was expressed as a hypothesis  
 \cite{srednicki-JPA96,srednicki1999approach},
 namely, eigenstate thermalization hypothesis (ETH).

 Technically, ETH is expressed as an ansatz for certain type of observable $O$ of a many-body system,
 stating that the matrix of $O$ on the energy basis of $\{ |E_i\ra \}$ exhibits a special structure
 and its elements are written in the following form,
 \begin{equation}\label{ETH}
 O_{ij} = O(E_i) \delta_{ij} +  f(E_i, E_j)r_{ij},
\end{equation}
 where $O(e)$ and $f(e,e')$ are smooth functions, 
 and $r_{ij} = r^*_{ji}$ are assumed to be independent random variables 
 with normal distribution (zero mean and unit variance).
 Originally, the function $f(e,e')$ was conjectured as being proportional to $e^{-S/2}$ 
 at the energy of $e_0=(e+e')/2$  for $e$ not far from $e'$ \cite{srednicki-JPA96},
 where $S$ is the thermodynamic entropy which
 is proportional to the particle number $N$ of the many-body system.
 From the perspective of quantum mechanics, 
 since the density of states $\rho_{\rm dos}$ increases exponentially with  $N$,
 $f(e,e')$ is sometimes written as proportional to $1/\sqrt{\rho_{\rm dos}}$.

 In recent years, the ETH ansatz has attracted lots of attention
 (see, e.g., Refs.\cite{ Rigol-AiP16,Deutch-RPP18,RS-PRL12,PSGC-PRL15} and references cited there).
 It is expected to supply a useful direction for the study of 
 thermalization of isolated many-body  quantum systems  (both for final results and for processes),
 a longstanding problem in physics
 which has attracted broad attentions recently
 \cite{PSSV-RMP11,EisertFG15-NP,Tasaki16-typi-therm,GE16-thermal-review,BISZ-PR16,Rigol-AiP16,Mori-IKU17-review}.
 The renewed interest was caused by progresses achieved in theoretical and experimental
 aspects, as well as improvements in the computation ability
 \cite{Deutch-RPP18,Rigol-AiP16,Bernien_2017,Turner_2018,Rigol_2008,
 Garrison_2018,Srednicki_Rigol_2013,Rigol_2021,Rigol_2020,Steinigeweg_2013,Rigol_2017,Vidmar_2019,Rigol_2019,Gemmer_2020}.

 Although lacking a rigorous analytical proof,
 vast numerical evidences have been found
 supporting validity of the ETH ansatz [Eq.(\ref{ETH})] in many models,
 even in models that are not completely chaotic according to spectral statistics
 (see, e.g., Refs.\cite{BMH-PRE14,BMH-PRE15,DLL-PRE18,Vidmar_2019,Vidmar21-PRB,YWW2022PS}
 and references cited in \cite{Rigol-AiP16,Deutch-RPP18}).
 The following behaviors of the function $f(e,e')$ have been found numerically
 with the increase of $|e-e'|$.
 That is, it usually shows a platform (scaling as $1/\sqrt{\rho_{\rm dos}}$) at $e$ sufficiently close $e'$, 
 followed by some power-law decay and then an exponential-type decay at $e$ sufficiently 
 far from $e'$.
 Local observables $O$ were considered in the original semiclassical arguments for ETH
 \cite{srednicki-JPA96,srednicki1999approach},
 while, later numerical simulations show that this requirement of locality seems too restrictive.
 However,  to make clear the scope of the observable $O$ for the ETH ansatz
 is a task too hard for numerical simulations, which is still an open problem.

 To get a satisfactory understanding of the ETH ansatz, 
 at least, the following five topics should be studied.

 (i) Analytical expression of the function $O(e)$ for the averaged diagonal elements.

 (ii) The following properties of the function $f(e,e')$ for the offdiagonal elements.
 That is, (a) an estimate to the height of the platform at small $|e - e'|$,
 including  its ${\rho_{\rm dos}^{-1/2}}$ scaling behavior;
 (b) an estimate to the width of the platform region, which may be related to 
 the relaxation time according to certain physical picture \cite{Jiaozi_2022,Serbyn_2017};
 (c) scope of the power-law decay and the decay exponent;
 and (d) exponent of the far exponential-type decay.

 (iii) Meaning of the stated randomness of the quantities $r_{ij}$.
 In fact, these quantities should not be completely random, 
 because energy eigenfunctions (EFs) of chaotic systems possess nonnegligible autocorrelation functions
 according to the semiclassical theory \cite{Berry77}. 
 Indeed,  nonnegligible correlations among $r_{ij}$
 were suggested by analytical arguments \cite{Dymarsky22BoundETH,Nussinov22,Vidmar20-prl}
 and were found in recent numerical simulations \cite{Jiaozi_2022,Dymarsky22BoundETH,Vidmar20-prl}.
 
  (iv) Relationship between $f(e,e')$ of $e\simeq e'$ for the offdiagonal elements
  and $f(e,e')$ of $e=e'$ for the diagonal elements.
  In fact, according to heuristic arguments \cite{PRA86-Feing-Peres} and 
  numerical simulations  \cite{YWW2022PS,Steinigeweg_2013,Rigol_2017,Gemmer_2020},
  the former may be about half of the latter. 
  
 And, (v) scope of validity of the ETH ansatz, particularly regarding the type of the system
 and the type of the operator $O$.

 In this paper, we are to study the first two topics discussed above.
 Regarding the function of $O(e)$,
 a semiclassical expression  was proposed previously for it  in the limit of $\hbar \to 0$ 
 \cite{PRA86-Feing-Peres}, based on physical arguments about ``reasonable'' operators,
 which have well-behaved classical limits.
 More rigorously, a similar result was obtained
 by sophisticated mathematical treatments (see references cited in Ref.\cite{srednicki-JPA96}). 
 We are to show that a semiclassical expression of $O(e)$ is derivable mathematically by a much simpler method, 
 which utilizes the concept of Weyl-ordered operator.
 And, even more, this method allows a direct 
 computation of quantum corrections to any order of $\hbar$ from the aspect of the observable $O$.

 Concerning the second topic discussed above, as is known,  
 derivation of a useful semiclassical expression of the function $f(e,e')$ 
 by making use of the original version of Berry's conjecture in the configuration space
 is not an easy task 
 \cite{srednicki-JPA96,srednicki1999approach}.
 In this paper, we use an alternative version of Berry's conjecture,
 which is expressed on an action basis  \cite{pre18-EF-BC},
 and study the possibility of deriving some useful expression by this approach.

 Moreover, one useful observation is that the above-discussed semiclassical treatments
  to the two functions of  $O(e)$ and $f(e,e')$
 do not rely on the many-body feature of the systems considered. 
 As a consequence, the analytical predictions may be tested 
 in models with a few degrees of freedom.
 Specifically, we are to employ the Lipkin-Meshkov-Glick (LMG) model \cite{Lipkin_1965} for this purpose.

 The paper is organized as follows. 
 In Sec.\ref{semi-EFs}, semiclassical understandings of EFs of quantum chaotic systems
 are recalled, which is basically given by Berry's conjecture. 
 In Sec.\ref{sect-semi- O(e)}, a semiclassical expression of $O(e)$ is derived 
 by making use of Weyl-ordered operators. 
 In Sec.\ref{sect-semi-f(e,e')}, a semiclassical study of $f(e,e')$
 is given by making use of Berry's conjecture on the action basis. 
 Numerical simulations in the LMG model 
 are given in Sec.\ref{sect-numerical}. 
 Finally, conclusions and discussions are given in Sec.\ref{sect-conclusion}.
 
\section{Semiclassical approach to EFs in quantum chaotic systems}\label{semi-EFs}
 
 In this section, we recall semiclassical descriptions of EFs,
 basically known as Berry's conjecture. 
 Specifically, some generic discussions on EFs are given in Sec.\ref{sect-generic}, 
 EFs in the configuration space are discussed in Sec.\ref{sect-EF-config},
 and those on the action basis in Sec.\ref{Berry_Conjecture_Action}. 
 
\subsection{Some generic discussions}\label{sect-generic}
 
 In the past half century or so, 
 a huge amount of knowledge has been accumulated in the field of quantum chaos
 (see, e.g., \cite{Casati-ed94,Haake-book}).
 The most-often-used criterion for quantum chaos was 
 proposed in the early 80s of the last century \cite{casati80,Bohigas84},
 which conjectures that the spectral statistics should be 
 in agreement with predictions of the random matrix theory.
 It then took more than twenty years to justify the above conjecture
 by making use of the semiclassical theory \cite{berry85,Sieber-Richter00,Sieber_2002,Haake04,Muller_Haake_2004}.
 Compared with spectral statistics, much less is known about statistical properties of energy
 eigenstates of chaotic systems, more exactly, of energy eigenfunctions (EFs).
 One major difficulty comes from the fact that EFs may behave quite differently on different bases.

 In most of the studies that have been carried out in EFs, 
 three types of basis were employed:
 (i) Position/momentum basis
	(see, e.g., Refs.\cite{Berry77,Robnik-LiBW,srednicki1994chaos});
 (ii) eigenbasis of some integrable (or regular) system, as an unperturbed basis
 (see, e.g., Refs.\cite{BISZ-PR16,Smilansky94,Meredith88,EFchaos-Benet03,pre18-EF-BC,pre17-EF-correlation,
 Wang-Robnik-Local20,Shapiro84});
 and (iii) eigenbasis of a nearby system (see, e.g., Refs.\cite{PRX20-Pandey-AGP,XLWW-CTP21}). 
 Below, we are to discuss EFs on bases of the former two types.

 For EFs in the configuration space (position basis), 
 one may distinguish between regions that are classically and energetically allowed and those  forbidden. 
 For EF components in the former regions, Berry's conjecture states that they may be regarded 
 as Gaussian random variables that possess certain correlations \cite{Berry77},
 the details of which are to be discussed in the next section (\ref{sect-EF-config}).
 For EF components in the latter regions, usually exponential behaviors are expected, 
 as well known in WKB treatment;
 meanwhile, certain perturbation approach may be useful in a discretized configuration space \cite{CPL_2004,CPL_2005}.

 Regarding bases of integrable systems, 
 the simplest one is given by eigenstates of the action. 
 Based on Berry's conjecture in the configuration space, 
 one may discuss EFs of quantum chaotic systems on an action basis.
 After being appropriately rescaled by the average shape of EFs, 
 the rescaled components of EFs usually obeys a Gaussian distribution \cite{pre18-EF-BC},
 as to be discussed in Sec.\ref{Berry_Conjecture_Action}. 
 While, a practically  useful expression for autocorrelation functions
 on this basis is still lacking.

\subsection{Chaotic EFs in the configuration space}\label{sect-EF-config}
 
 In this section, we recall basic contents of Berry's conjecture. 
 Consider a quantum system, 
 which possesses an effective Planck constant $\hbar$,
 such that a classical-counterpart system in a $f$-dimensional configuration space
 may be obtained in the limit of $\hbar \to 0$. 
 Coordinates in the classical phase space are indicated as $(\bp, \bq)$, 
 where $\bm{p}=(p_1,p_2,\cdots, p_f)$ and $\bm{q}=(q_1,q_2,\cdots, q_f)$.
 The classical counterpart system is assumed chaotic with a positive 
 maximum Lyapunov exponent.

 The Hamiltonian of the system is denoted by $H$
 as a function of $\hat\bp$ and $\hat\bq$, namely, $H(\hat\bp,\hat\bq)$.
 \footnote{ For the sake of clearness in discussion,
 hats are written for the operators of $\hat \bp$ and $\hat \bq$
 (also for $\hat{\boldsymbol{I}}$ of action).}
 Its eigenenergies are written as $E_i$,
 in the increasing energy order, and eigenstates are indicated as $|E_i\ra$, 
\begin{align}
 H|E_i\rangle = E_i|E_i\rangle.
\end{align}
 The EF of $|E_i\ra$ in the configuration space is indicated as $\psi_i(\bq)$,
 $\psi_i(\bq)= \la \bq |E_i\ra$.

 The Wigner distribution function corresponding to $\psi_i(\bq)$,
 denoted by $W_i(\bp,\bq)$, is written as
\begin{align}\label{}
W_i(\bp,\bq)=\frac{1}{(2\pi\hbar)^{f}}\int_{-\infty}^{\infty}\psi_i^{*}
(\bq+\frac{\br }{2})\psi_i(\bq-\frac{\br}{2})
e^{i\bp\cdot \br/\hbar}d\br.
\end{align}
 The average shape of EFs within  a narrow energy shell centered at $E_i$
 is denoted by $\Pi_i(\bq)$,
\begin{align}\label{}
 \Pi_i(\bq) = \overline{|\psi_i(\bq)|^2}.
\end{align}
 Here and hereafter, an overline indicates average taken over an energy shell of the
 (perturbed) system $H$. 
 \footnote{ In the original work of Berry, the average was taken over 
 a small region in the configuration space \cite{Berry77}. 
 Later, it was argued that a better method is to take average over a narrow energy shell
 \cite{srednicki1994chaos,Berry_1989}. 
 }
 More exactly, for a narrow energy window $\epsilon$,
 one may consider a coarse-grained $\delta$-function, indicated as $\delta_\epsilon (e,E_i)$,
\be
\delta_{\epsilon}(e,E_i)=\begin{cases}
\frac{1}{\epsilon} & e\in[E_i -\frac{\epsilon}{2}, E_i +\frac{\epsilon}{2}],\\
0 & {\rm otherwise},
\end{cases}
\ee
 and take average within this window. 
 The energy window $\epsilon$ should 
 be small in the classical case, such that change of the energy surface is negligible
 within the window;
 meanwhile, it be sufficiently large in the quantum case, such that many energy levels
 are included within the window.

 The average shape of EF may be computed from the averaged Wigner function, 
\begin{align}\label{}
 \Pi_i(\bq) = \int d\bp \ov W_i(\bp,\bq). 
\end{align}
 Based on semiclassical arguments, it was conjectured that the averaged Wigner function 
 is approximately given by the corresponding classical energy surface in phase space
  \cite{Berry77,Berry_1991,Voros_1976,Voros_1977,pre18-EF-BC}, i.e., 
\be\label{eq-WFchaos}
\overline{W}_i (\boldsymbol{p},\boldsymbol{q})
 \simeq \frac{\delta(H_{\rm cl}(\boldsymbol{p},\boldsymbol{q})-E_i)}{S({E_i})},
\ee
 where $S(E)$ represents the area of the energy surface with $H_{\rm cl}(\boldsymbol{p},\boldsymbol{q})=E$,
\be \label{SE}
S({E})=\int d\boldsymbol{p}d\boldsymbol{q}\delta(E-H_{\rm cl}(\boldsymbol{p},\boldsymbol{q})).
\ee
 Here and hereafter, for an arbitrary operator function $O(\hat\bp,\hat\bq)$,
 $O_{\rm cl}(\bp,\bq)$ with a subscript ``cl'' 
 refers to the classical quantity, which is obtained by directly replacing $(\hat\bp, \hat\bq)$
 in $O(\hat\bp,\hat\bq)$ by the classical coordinates $(\bp, \bq)$.

 By definition, the two-point autocorrelation function is written as
\begin{align}\label{Cq}
  C(\bq, \bq') \equiv C(\br, \bq_0) = \frac{1}{\Pi_i(\bq_0)} \ov{ \psi_i^* (\bq_0+\frac{\br }{2})\psi_i(\bq_0-\frac{\br}{2})},
\end{align}
 where
\begin{align}\label{q0-r}
 \bq_0 = \frac{\bq + \bq'}{2}, \quad \br = \bq' - \bq.
\end{align}
 It has the following relation to the Wigner function, 
\begin{align}\label{Crq0-ovW}
 C(\br, \bq_0) = \frac{1}{\Pi_i(\bq_0)} \int d\bp \ov W_i(\bp,\bq_0)e^{-i\bp\cdot \br/\hbar}.
\end{align}
 Then, one sees that
 \begin{equation}
  C(\bm{r},\bm{q}_0) \simeq \frac{\int d\bm{p}\ \delta\left[H_{\rm cl}(\bm{p},\bm{q}_0)
  -E_i\right]e^{-i\bm{p}\cdot\bm{r}/\hbar}}{\int d\bm{p}\ \delta\left[H_{\rm cl}(\bm{p},\bm{q}_0)-E_i\right]}.
 \end{equation}
 According to Berry's conjecture \cite{Berry77}, 
 the EF $\psi$ is regarded as a Gaussian random function of $\bq$, whose statistical properties 
 (probability distributions of $\psi$ and of its derivatives, correlations between $\psi$ at
 two or more points, etc.) are all determined by $\Pi(\bq)$ and $C(\br,\bq_0)$,
 which are computed from the averaged Wigner distribution function given above.

\subsection{EFs on the action basis}\label{Berry_Conjecture_Action}
 
 In this section, we recall a version of Berry's conjecture, which is 
 expressed on an action basis \cite{pre18-EF-BC}.
 The perturbed Hamiltonian is written as
\begin{equation}\label{H}
H=H_0+\lambda V,
\end{equation}
 where $H_0$ indicates the Hamiltonian of an integrable system and $V$ is a generic perturbation,
 with a parameter $\lambda$ for adjusting the perturbation strength.
 In terms of operators for the action, $H_0$ is written as
\be
H_0=\boldsymbol{d}\cdot \hat{\boldsymbol{I}} +c_0,
\ee
 where $\hat{\boldsymbol{I}}=(\hat I_1, \hat I_2,\cdots, \hat I_f)$ indicate the action operators
 \footnote{More rigorously to say, $\hat{\boldsymbol{I}}$ are those operators
 that correspond to the action variables in the classical counterpart of the system $H_0$.},
 $\boldsymbol{d }$ is a parameter vector, $\boldsymbol{d }=(d _{1},d _{2},\cdots,d _{f})$, 
 and $c_0$ is a constant parameter.

 We use $|\boldsymbol n\ra$ to denote the eigenbasis of operator vector $\hat{\bm{I}}$,
 with $\hat{\bm{I}} |\boldsymbol n\ra = \boldsymbol {I_n}|\boldsymbol n\ra$,
 where $\boldsymbol{n}=(n_1,n_2,\cdots,n_f)$ is an integer vector and $\boldsymbol{I_n}=\boldsymbol n \hbar$.
 The Hamiltonian $H_0$ has a diagonal form on the action basis,
 with eigenvalues denoted by $E_{\boldsymbol{n}}^{0}$,
\begin{equation}
H_{0}|\boldsymbol{n}\rangle=E_{\boldsymbol{n}}^{0}|\boldsymbol{n}\rangle.
\end{equation}
 The expansion of $|E_i \ra$ on the basis $|\boldsymbol{n}\rangle$ is written as
\be
|E_i\rangle=\sum_{\boldsymbol{n}}C^{i}_{\bs{n}}|\boldsymbol{n}\rangle,
\ee
 where $C^{i}_{\bs{n}} = \la \boldsymbol{n}|E_i \ra$.
 We use $\psi_{\boldsymbol{n}}^{0}(\bq )$ to denote 
 the wave function of $|\boldsymbol{n}\rangle$ in the configuration space.
 Then, the components $C^{i}_{\boldsymbol{n}}$ are written as
\be\label{Calpha-int}
C^{i}_{\bs{n}}=\int \left( \psi_{\bs{n}}^{0}(\bq ) \right)^{*}\psi_{i }(\bq )d\bq .
\ee

 One may make use of the following well-known expression of $|C^{i}_{\bs{n}}|^{2}$,
\begin{equation}
|C^{i}_{\bs{n}}|^{2}=(2\pi\hbar)^{f}\int d\boldsymbol{p}d\boldsymbol{q}W_{i }
(\boldsymbol{p},\boldsymbol{q})W_{\boldsymbol{n}}^{0}(\boldsymbol{p},\boldsymbol{q}),
\label{eq-wigner-wave}
\end{equation}
 where $ W_{\bm{n}}^{0}(\bm{p},\bm{q})$ is the Wigner function of $\psi_{\boldsymbol{n}}^{0}(\bq )$,
 and find that \cite{pre18-EF-BC}
\begin{equation}\label{C_Average_Cell_Cell}
  \begin{aligned}
  \overline{|C^{i}_{\bm{n}}|^{2}}
    =&(2\pi\hbar)^{f}\int d\bm{p}d\bm{q} \ov W_i(\bm{p},\bm{q}) \la W_{\bm{n}}^{0}(\bm{p},\bm{q}) \ra.
  \end{aligned}
\end{equation}
 Here, $\la \cdot \ra$ indicates an average over
 a small region in the phase space (or configuration space) for the integrable (unperturbed) system. 
 \footnote{For the Wigner function of an integrable system, 
 the average should not be taken over a narrow energy shell
 of the unperturbed energy $E_{\boldsymbol{n}}^{0}$, since 
 values of the action number $\bm{n}$ of nearby levels are usually far from each other. }
 It was conjectured that \cite{Berry77}
  \begin{align}\label{Wigner-integrable}
    \la W_{\bm{n}}^{0}(\bm{p},\bm{q}) \ra & \simeq 
    \frac{\delta^f\left[\bm{I}(\bm{p},\bm{q})-\bm{I}_{\bm{n}}\right]}{(2\pi)^f}.
  \end{align}

Then,  one finds that 
\be\label{eq-EFshapecl}
\ov {|C^{i}_{\bs{n}}|^{2}} \simeq \hbar^{f}\ \Pi(E_i,\boldsymbol{I_{n}}),
\ee
 where
  \begin{gather}\label{Pi-E-I}
    \Pi(E,\boldsymbol{I}) = \frac{S(E,\boldsymbol{I})}{S({E})}.
\end{gather}
 Here, $S(E,\boldsymbol{I})$ represents the overlap of the energy
 surface of $H_{\rm cl}(\boldsymbol{p},
\boldsymbol{q}) =E$ and the torus of $\boldsymbol{I}_{\rm cl}(\boldsymbol{p},\boldsymbol{q}) =\boldsymbol{I}$,
  \begin{gather}
 \label{eq-ov1}
    S(E,\boldsymbol{I})=\int d\boldsymbol{p}d\boldsymbol{q}\delta(E-H_{\rm cl}(\boldsymbol{p},
    \boldsymbol{q}))\delta^f(\boldsymbol{I}-\boldsymbol{I}_{\rm cl}(\boldsymbol{p},\boldsymbol{q})).
\end{gather}
 Since Eq.(\ref{eq-WFchaos}) works for classically allowed regions only,
 so does Eq.(\ref{eq-EFshapecl}).
 Sometimes, quantities like $\Pi(E,\boldsymbol{I})$ are referred to as
 {classical analog} of averaged EFs \cite{EFchaos-Benet03,EFchaos-Benet00,EFchaos-Borgonovi98}.

 According to Berry's conjecture, the rescaled components denoted by $R^i_{ \boldsymbol{n}}$,
\be\label{eq-C9}
 R^i_{ \boldsymbol{n}}=\frac{C^{i}_{\bs{n}}}{\sqrt{\ov{\abs{C^{i}_{\bs{n}}}^{2}}}},
\ee
 should have a Gaussian distribution,
 which has been confirmed by numerical simulations \cite{pre18-EF-BC}.
 The autocorrelation function on the action basis is given by the average of $C^{i*}_{{\bs n}'} C^i_{\bs n}$
 over a narrow energy window around $E_i$,
\begin{gather}\label{corre-action-basis}
 \ov{C^{i*}_{{\bs n}'} C^i_{\bs n}} = \int d\bq'  d\bq 
  \psi_{\bs{n}'}^{0}(\bq' )  \psi_{\bs{n}}^{0*}(\bq ) \ov{\psi^*_{i }(\bq' ) \psi_{i }(\bq )}
\end{gather}
 Due to the nonvanishing autocorrelation function $\ov{\psi^*_{i }(\bq' ) \psi_{i }(\bq )}$ in the configuration space, 
 the EFs on the action basis should posses nonvanishing autocorrelation functions, too.
 That is, $\ov{C^{i*}_{{\bs n}'} C^i_{\bs n}}$ should be nonvanishing and the same for 
 $\ov{R^{i*}_{{\bs n}'} R^i_{\bs n}}$,
 though it is not an easy task to derive a useful expression for them.

\section{The digonal function $O(e)$}\label{sect-semi- O(e)}

 In this section, we discuss a method of deriving 
 semiclassical expressions of the function $O(e)$ for averaged diagonal elements in the ETH ansatz.
 This method is valid not only in the limit of $\hbar \to 0$, 
 and also useful for getting finite-$\hbar $ contributions.
 The basic idea of the method is simple, which is to write the studied operator 
 as a sum of Weyl-ordered operators.

 Specifically, generic discussions are given in Sec.\ref{sect-Oe-generic}. 
 The case of one-dimensional configuration space is discussed in Sec.\ref{sect-Oe-Weyl-f1}
 for an operator $O$ as a Weyl-ordered operator 
 and in Sec.\ref{sect-Oe-gen-f1} for $O$ as a generic polynomial function of position and momentum. 
 Finally, generalization of the above results to 
 the case of a multi-dimensional configuration space is discussed in Sec.\ref{sect-Oe-gen}.

\subsection{Generic discussions}\label{sect-Oe-generic}

 Semiclassical studies of $O(e)$ 
 were carried out by several groups of authors in the last century, getting similar results 
 in the limit of $\hbar \to 0$.
 Specifically, based on sophisticated mathematical treatments, 
 in the limit of $\hbar \to 0$, the following semiclassical expression 
 was given for chaotic eigenfunctions in the high-energy region  (see references cited in Ref.\cite{srednicki-JPA96}), 
 i.e., 
\begin{gather}\label{O(E)-sc}
 O(E_i) \simeq \frac 1{S(E_i)}
 {\int d\bp d\bq \delta (H_{\rm cl}(\bp,\bq)- E_i) O_{\rm cl}(\bp,\bq)}.
\end{gather}
 Meanwhile, the same semiclassical expression was given
 in Ref.\cite{PRA86-Feing-Peres} mainly based on physical argument
 for those ``reasonable'' operators $O$ that have well behaved classical limits as $\hbar \to 0$.

 For the purpose of studying the function of $O(e)$,
 one may write $O_{ii} = \la E_i|O(\hat{\bm{p}},\hat{\bm{q}})|E_i\ra$ as follows
 by inserting identity operators, 
\begin{align}\label{}
 & O_{ii} 
 = \int d\bq' d\bq \psi^*_i(\bq') \psi_i(\bq) O(\bq',\bq),
\end{align}
 where 
\begin{align}\label{<q'|O|q>}
 O(\bq',\bq) = \la \bq'|O(\hat{\bm{p}},\hat{\bm{q}})|\bq \ra. 
\end{align}
 Taking average over a narrow energy shell centered at $E_i$
 and making use of Eqs.(\ref{Cq})-(\ref{Crq0-ovW}), one gets that
\begin{gather}\label{Oii-1}
 O(E_i)  = \int d\bq_0  d\bp \ov W_i(\bp,\bq_0)  \ww O(\bp,\bq_0), 
\end{gather}
 where
\begin{align}\label{wwO}
 \ww O(\bp,\bq_0) = \int d\br e^{-i\bp\cdot \br/\hbar} O(\bq',\bq),
\end{align}
 with $\bq_0$ and $\br$ defined in Eq.(\ref{q0-r}).  
 One sees that,
 with the averaged Wigner function approximated by Eq.(\ref{eq-WFchaos}), 
 the function $O(e)$ is basically determined by the function of $\ww O(\bp,\bq_0)$.

 As mentioned previously, we are to employ an alternative method in the study of the function $O(e)$,
 which makes use of the concept of Weyl-ordered operator. 
 One key observation is that, if the operator $O$ is Weyl-ordered,
 then, $\ww O(\bp,\bq_0)$ turns out to have the same form as the classical function $O_{\rm cl}$;
 and, furthermore, this treatment is generalizable to a generic polynomial operator $O$. 
 For the sake of simplicity in discussion, 
 in the following sections, we first discuss the function of $\ww O$ in a one-dimensional
 configuration space and, then, in a multi-dimensional configuration space. 

\subsection{$O(e)$ as a Weyl-ordered operator at $f=1$}\label{sect-Oe-Weyl-f1}

 In this section, we discuss Weyl-ordered operator in a $1$-dimensional configuration space.
 As is known, an operator is said to be Weyl-ordered,  indicated as $A^{\rm Weyl}$, if it 
 has the following binomial form \cite{Peskin_2019},
\begin{equation}\label{WO-main}
  A^{\rm Weyl}(\hat{p},\hat{q})=\sum_{n=0}^\infty \frac{1}{2^n}
  \sum_{k=0}^n \binom{n}{k}\hat{q}^{n-k}f_n(\hat{p})\hat{q}^k,
\end{equation}
 where $f_n(\hat{p})$ is some function of $\hat{p}$, which is $k$-independent while may 
 be $n$-dependent, and $\binom{n}{k}$ is the combination number,
\begin{equation}
  \binom{n}{k} =\frac{n!}{k!(n-k)!}.
\end{equation}
 It is direct to check that, at each fixed value of $n$, the term on the right-hand side (rhs) of Eq.(\ref{WO-main}) 
 has the following property, 
\begin{equation}\label{Ocl2Oqq'-main}
  \begin{aligned}
 &    \bra{q+\frac{r}{2}} \frac{1}{2^n}
  \sum_{k=0}^n \binom{n}{k}\hat{q}^{n-k}f_n(\hat{p})\hat{q}^k \ket{q-\frac{r}{2}}
  \\ &  = q^n \bra{q+\frac{r}{2}}f_n(\hat{p})\ket{q-\frac{r}{2}}.
  \end{aligned}
\end{equation}

 Inserting the identity operator of $\int dp \ketbra{p}{p}$ before 
 the term $f_n(\hat{p})$ on the rhs of Eq.(\ref{Ocl2Oqq'-main})
 and then making use of Eq.(\ref{WO-main}),  one gets that
\begin{equation}\label{A_WeylOperator2A_Weylcl}
  \begin{aligned}
    \bra{q+\frac{r}{2}}A^{\rm Weyl}(\hat{p},\hat{q})&\ket{q-\frac{r}{2}}
    =\int dp\ \frac{e^{i p \cdot r/\hbar}}{2\pi\hbar}  \sum_{n=0}^\infty q^n f_n(p)\\
    =&\frac{1}{2\pi\hbar}\int dp\ e^{i p \cdot r/\hbar}A^{\rm Weyl}_{\rm cl}(p,q).
  \end{aligned}
\end{equation}
 Here,  
\begin{align}\label{O-Weyl-cl}
 A^{\rm Weyl}_{\rm cl}(p,q) = \sum_{n=0}^\infty q^{n} f_n(p),
\end{align}
 which is gotten by directly replacing the operators $(\hat{q},\hat{p})$ on the rhs of 
 Eq.(\ref{WO-main}) by the classical variables $(q,p)$.
 Then, for $O = A^{\rm Weyl}$ in Eq.(\ref{wwO}), making use of Eqs.(\ref{<q'|O|q>}) and 
 (\ref{A_WeylOperator2A_Weylcl}), one finds that
\begin{equation}\label{Opq_Classical}
\widetilde{A}^{\rm Weyl}(p,q)=A^{\rm Weyl}_{\rm cl}(p,q).
\end{equation}
 According to Eq.(\ref{Oii-1}), this implies that
\begin{gather}\label{}
 A^{\rm Weyl}(E_i)  = \int dp dq_0  \ov W_i(p,q_0)  A^{\rm Weyl}_{\rm cl}\left({p},{q_0}\right), \label{Oii-2}
\end{gather}
 which gives Eq.(\ref{O(E)-sc}) under the semiclassical approximation 
 of the Wigner function in Eq.(\ref{eq-WFchaos}). 

\subsection{$O(e)$ as a generic operator at $f=1$}\label{sect-Oe-gen-f1}

 Now, we discuss an operator $O$ as a generic polynomial function of $\hat p$ and $\hat q$. 
 We use $P_{m,n}^{(s)}(\hat{p},\hat{q})$ to indicate one product term in this polynomial function, 
 where $m$ is the number of $\hat{p}$ contained in the product, $n$ is the number of $\hat{q}$, 
 and $s$ is a label representing the order of $\hat{p}$ and $\hat{q}$ in the product. 
 For example, $P_{2,1}^{(s_0)}(\hat{p},\hat{q})=\hat{q}\hat{p}^2$,
 $P_{2,1}^{(s_1)}(\hat{p},\hat{q})=\hat{p}\hat{q}\hat{p}$,
 $P_{1,1}^{(s_2)}(\hat{p},\hat{q})=\hat{q}\hat{p}$, etc.
 Thus, the operator $O$ is written as
\begin{align}\label{O-generic-operator}
  O(\hat{p},\hat{q})
  = \sum_{m,n}  \sum_s C_{m,n}^{(s)} P_{m,n}^{(s)}(\hat{p},\hat{q}),
\end{align}
 where $C_{m,n}^{(s)}$ are coefficients. 
 Clearly, the classical quantity $O_{\rm cl}(p,q)$ is written as
\begin{align}\label{Ocl}
O_{\rm cl}(p,q) =   \sum_{m,n} p^m q^n \left( \sum_s C_{m,n}^{(s)} \right).
\end{align}

 One key point is that each product $P_{m,n}^{(s)}(\hat{p},\hat{q})$
 may be written as a sum of Weyl-ordered products. 
 As an example, let us consider $P_{2,2}^{(s)}(\hat{p},\hat{q})=\hat{q}\hat{p}\hat{q}\hat{p}$.
 One easily checks that
\begin{equation}\label{WeylOrder_Example}
  \begin{aligned}
    &P_{2,2}^{(s)}(\hat{p},\hat{q})
    =\frac{1}{4}\left(\hat{q}\hat{p}\hat{q}\hat{p}+2\hat{q}\hat{p}\hat{q}\hat{p}+\hat{q}\hat{p}\hat{q}\hat{p}\right)\\
    =&\frac{1}{4}\left(\hat{q}^2\hat{p}^2+2\hat{q}\hat{p}^2\hat{q}+\hat{p}^2\hat{q}^2\right)
    +\frac{1}{4}i\hbar\left(2\hat{p}\hat{q}+\hat{q}\hat{p}\right)\\
    =&\frac{1}{4}\left(\hat{q}^2\hat{p}^2+2\hat{q}\hat{p}^2\hat{q}+\hat{p}^2\hat{q}^2\right)
    +\frac{3}{8}i\hbar\left(\hat{q}\hat{p}+\hat{p}\hat{q}\right)+\frac{1}{8}\hbar^2.\\
  \end{aligned}
\end{equation}
 In the notation to be introduced below, it is written as
\begin{equation}
  P_{2,2}^{(s_0)}(\hat{p},\hat{q}) 
  =P^{\rm Weyl}_{2,2}(\hat{p},\hat{q})+Q^{(s_0) }_{2,2}(\hat{p},\hat{q}),
\end{equation}
where
\begin{equation}
    Q^{(s_0) }_{2,2}(\hat{p},\hat{q})
    =\frac{3}{4}i\hbar P^{\rm Weyl}_{1,1}(\hat{p},\hat{q})-\frac{1}{8}(i\hbar)^2P^{\rm Weyl}_{0,0}(\hat{p},\hat{q}).
\end{equation}

 For a generic product $P_{m,n}^{(s)}(\hat{p},\hat{q})$,
 after some derivation, one finds the following expression (see Appendix \ref{App_P_Weyl_Modify} for a proof),
\begin{equation}\label{P_Weyl_Modify}
  P_{m,n}^{(s)}(\hat{p},\hat{q})=P^{\rm Weyl}_{m,n}(\hat{p},\hat{q})+Q^{(s) }_{m,n}(\hat{p},\hat{q}).
\end{equation}
 Here, $P^{\rm Weyl}_{m,n}(\hat{p},\hat{q})$ indicates a {Weyl-ordered} operator
 which corresponds to the two indices of $(m,n)$, i.e., 
\begin{align}
    P^{\rm Weyl}_{m,n}(\hat{p},\hat{q}) =&\frac{1}{2^n}\sum_{k=0}^n 
    \binom{n}{k}\hat{q}^{n-k}\hat{p}^m\hat{q}^k;\label{P_WeylOrder}
\end{align}
 and, $Q^{(s) }_{m,n}(\hat{p},\hat{q})$ is a sum of Weyl-ordered operators as given below, 
\begin{align}
    Q^{(s) }_{m,n}(\hat{p},\hat{q}) =&\sum_{\lambda=1}^{\min(m,n)} 
    (i\hbar)^\lambda D_{m-\lambda,n-\lambda}^{(s)} P^{\rm Weyl}_{m-\lambda,n-\lambda}(\hat{p},\hat{q})
    \label{P_Remainder},
  \end{align}
 where $D_{m-\lambda,n-\lambda}^{(s)}$ represent certain coefficients
 [see Eq.(\ref{D_Coefficient}) in Appendix \ref{App_P_Weyl_Modify} for its exact expression].
 Note that the label $s$ in $Q^{(s) }_{m,n}(\hat{p},\hat{q})$ merely indicates that this term comes from 
 the product of $P_{m,n}^{(s)}(\hat{p},\hat{q})$ (similar for $D_{m-\lambda,n-\lambda}^{(s)}$).
 Replacing $(\hat{p},\hat{q})$ by $(p,q)$, it is easy to check that
\begin{equation}
  \left(P^{(s)}_{m,n}\right)_{\rm cl}(p,q)=\left(P^{\rm Weyl}_{m,n}\right)_{\rm cl}(p,q)=p^m q^n,
\end{equation}
 independent of the label $s$.

 Inserting Eq.(\ref{P_Weyl_Modify}) into Eq.(\ref{O-generic-operator}), 
 one gets the following linear expansion of $O(\hat{p},\hat{q})$,
 as a sum of Weyl-ordered operators,
\begin{equation}\label{O_Weyl_Modify}
  \begin{aligned}
    O(\hat{p},\hat{q})=&\sum_{m,n} P^{\rm Weyl}_{m,n}(\hat{p},\hat{q}) \left( \sum_s C_{m,n}^{(s)} \right) \\
    &+\sum_{m,n}\sum_s C_{m,n}^{(s)}Q^{(s) }_{m,n}(\hat{p},\hat{q}).
  \end{aligned}
\end{equation}
 Substituting the above expression into Eq.(\ref{<q'|O|q>}) and then into Eq.(\ref{wwO}), 
 and making use of Eqs.(\ref{Opq_Classical}) and (\ref{Ocl}), one finds that
\begin{align}\label{ww-Q-cl}
 \ww O(p,q) = O_{\rm sc}(p,q),
\end{align}
 where $O_{\rm sc}(p,q)$ is a semiclassical function obtained from
 the operator function of $O$ by the following method, 
\begin{equation}\label{Osc_Definition}
  \begin{aligned}
   & O_{\rm sc}(p,q)  :  = O_{\rm cl}(p,q)
    \\ & + \sum_{m,n}\sum_{\lambda=1}^{\min(m,n)} (i\hbar)^\lambda p^{m-\lambda}q^{n-\lambda}
   \left( \sum_s  C_{m,n}^{(s)}  D_{m-\lambda,n-\lambda}^{(s)} \right).
  \end{aligned}
\end{equation}
 Then, from Eq.(\ref{Oii-1}), one gets that
\begin{equation}\label{OEi_Semiclassical}
  O(E_i)=\int dp dq_0 \overline{W}_i(p,q_0)O_{\rm sc}(p,q_0).
\end{equation}

\subsection{$O(e)$ for a multi-dimensional configuration space}\label{sect-Oe-gen}

 The above discussions are generalizable to operators $O$ in a multi-dimensional configuration space.
 We use $\mu$ of $\mu =1,\ldots, f$ to label the degrees of freedom, e.g.,
$\bq = (q_1, \ldots, q_\mu, \ldots, q_f)$.

 In this generic case, the concept of Weyl-ordered operator is a direct generalization of 
 that given previously in the one-dimensional case, by means of direct product. 
 More exactly, a Weyl-ordered operator is written as follows, 
\begin{equation}\label{A-Weyl-multi}
  A^{\rm Weyl}(\hat{\bm{p}},\hat{\bm{q}})=\prod_{\mu=1}^{f}A^{\rm Weyl}_\mu(\hat{p}_\mu,\hat{q}_\mu),
\end{equation}
 where $A^{\rm Weyl}_\mu(\hat{p}_\mu,\hat{q}_\mu)$ is a Weyl-ordered operator
 defined in Eq.(\ref{WO-main}) for the $\mu$-th degree of freedom. 
 Note that
\begin{equation}\label{HighOrder_pqCommucate}
  [\hat{p}^\mu,\hat{q}^\nu]=[\hat{p}^\mu,\hat{p}^\nu]=[\hat{q}^\mu,\hat{q}^\nu]=0
  \quad \text{for $\mu\neq\nu$},
\end{equation}
 and, hence, the order of $A^{\rm Weyl}_\mu(\hat{p}_\mu,\hat{q}_\mu)$ on the 
 rhs of Eq.(\ref{A-Weyl-multi}) is irrelevant. 
 One sees that
\begin{equation}
  \bra{\bm{q}'}A^{\rm Weyl}(\hat{\bm{p}},\hat{\bm{q}})\ket{\bm{q}}
  =\prod_{\mu=1}^{f}\bra{q'_\mu}A^{\rm Weyl}_\mu(\hat{p}_\mu,\hat{q}_\mu)\ket{q_\mu}.
\end{equation}
 Making use of Eq.(\ref{A_WeylOperator2A_Weylcl}) for each degree of freedom, it is easy to find that
\begin{equation}\label{A_WeylO2A_Weylcl_HighDim}
  \begin{aligned}
    \bra{\bm{q}+\frac{\bm{r}}{2}}A^{\rm Weyl}(\hat{\bm{p}},&\hat{\bm{q}})\ket{\bm{q}-\frac{\bm{r}}{2}}\\
    =&\frac{1}{(2\pi\hbar)^f}\int d\bm{p}\ e^{i \bm{p} \cdot \bm{r}/\hbar}A^{\rm Weyl}_{\rm cl}(\bm{p},\bm{q}),
  \end{aligned}
\end{equation} 
where
\begin{equation}
  A^{\rm Weyl}_{\rm cl}(\bm{p},\bm{q})=\prod_{\mu=1}^{f}A^{\rm Weyl}_{\mu,\rm cl}(p_\mu,q_\mu),
\end{equation}
 with $A^{\rm Weyl}_{\mu,\rm cl}(p_\mu,q_\mu)$ given by Eq.(\ref{O-Weyl-cl}).

 An operator $O(\hat{\bm{p}},\hat{\bm{q}})$, which is a generic polynomial 
 function of $\hat{\bm{p}}$ and $\hat{\bm{q}}$, is written in a form similar to the one-dimensional 
 case in Eq.(\ref{O-generic-operator}).
 That is,
\begin{equation}\label{O_HighDim}
  O(\hat{\bm{p}},\hat{\bm{q}})=\sum_{\bm{m},\bm{n}}  \sum_{\bm{s}} C_{\bm{m},\bm{n}}^{(\bm{s})} 
  P_{\bm{m},\bm{n}}^{(\bm{s})}(\hat{\bm{p}},\hat{\bm{q}}),
\end{equation}
 where  $\bm{m} =(m_1,\cdots,m_f)$, $\bm{n} =(n_1,\cdots,n_f)$, $\bm{s} =(s_1,\cdots,s_f)$,
 $C_{\bm{m},\bm{n}}^{(\bm{s})}$ are coefficients, and 
\begin{equation}\label{Pmns-generic}
  P_{\bm{m},\bm{n}}^{(\bm{s})}(\hat{\bm{p}},\hat{\bm{q}}) =
  \prod_{\mu=1}^{f}P^{(s_\mu)}_{m_\mu,n_\mu}(\hat{p}_\mu,\hat{q}_\mu).
\end{equation}
 Here, $P^{(s_\mu)}_{m_\mu,n_\mu}(\hat{p}_\mu,\hat{q}_\mu)$ are $1$-dimensional products
 like those discussed in the above section. 
 Substituting Eq.(\ref{P_Weyl_Modify}) into Eq.(\ref{Pmns-generic}), one gets that
\begin{equation}\label{P_Weyl_Modify_HighDim}
  P_{\bm{m},\bm{n}}^{(\bm{s})}(\hat{\bm{p}},\hat{\bm{q}})
  =P^{\rm Weyl}_{\bm{m},\bm{n}}(\hat{\bm{p}},\hat{\bm{q}})
  +Q^{(\bm{s}) }_{\bm{m},\bm{n}}(\hat{\bm{p}},\hat{\bm{q}}),
\end{equation}
where
\begin{subequations}\label{P_Weyl&Remainder_HighDim}
  \begin{align}
    &P_{\bm{m},\bm{n}}^{\rm Weyl}(\hat{\bm{p}},\hat{\bm{q}})
    =\prod_{\mu=1}^{f}P^{\rm Weyl}_{m_\mu,n_\mu}(\hat{p}_\mu,\hat{q}_\mu),\label{P_WeylOrder_HighDim}\\
\notag    &Q^{(\bm{s}) }_{\bm{m},\bm{n}}(\hat{\bm{p}},\hat{\bm{q}}):=
    \prod_{\mu=1}^{f}\left[P^{\rm Weyl}_{m_\mu,n_\mu}(\hat{p}_\mu,\hat{q}_\mu)
    +Q^{(s_\mu)}_{m_\mu,n_\mu}(\hat{p}_\mu,\hat{q}_\mu)\right]
    \\ & \qquad \qquad -P_{\bm{m},\bm{n}}^{\rm Weyl}(\hat{\bm{p}},\hat{\bm{q}})
    \label{P_Remainder_HighDim}
  \end{align}
\end{subequations}
 Note that $Q^{(\bm{s}) }_{\bm{m},\bm{n}}(\hat{\bm{p}},\hat{\bm{q}})$
 is also a sum of Weyl-ordered products. 
 Finally, substituting Eq.(\ref{P_Weyl_Modify_HighDim}) into Eq.(\ref{O_HighDim}), one 
 writes the operator $O(\hat{\bm{p}},\hat{\bm{q}})$ as a sum of Weyl-ordered operators,
\begin{equation}\label{O_Weyl_Modify_HighDim}
  \begin{aligned}
    O(\hat{\bm{p}},\hat{\bm{q}})=&\sum_{\bm{m},\bm{n}} P^{\rm Weyl}_{\bm{m},\bm{n}}(\hat{\bm{p}},\hat{\bm{q}})\sum_{\bm{s}} C_{\bm{m},\bm{n}}^{(\bm{s})}\\
    &+\sum_{\bm{m},\bm{n}}\sum_{\bm{s}} C_{\bm{m},\bm{n}}^{(\bm{s})}Q^{(\bm{s}) }_{\bm{m},\bm{n}}(\hat{\bm{p}},\hat{\bm{q}}).
  \end{aligned}
\end{equation}

 Similar to Eq.(\ref{ww-Q-cl}) in the $1$-dimensional case, now, one gets that
\begin{align}\label{ww-Q-cl-g}
 \ww O(\bp,\bq) = O_{\rm sc}(\bp,\bq),
\end{align}
 where $O_{\rm sc}(\bp,\bq)$ is a semiclassical function,
\begin{equation}\label{Osc_Definition_HighDim}
  \begin{aligned}
    O_{\rm sc}(\bm{q},\bm{p})=O_{\rm cl}(\bm{q},\bm{p})+O'_{sc}(\bm{q},\bm{p}).
  \end{aligned}
\end{equation}
 Here, 
\begin{subequations}
  \begin{align}
    &O_{\rm cl}(\bm{q},\bm{p})=\sum_{\bm{m},\bm{n}} \prod_{\mu=1}^{f}p_\mu^{m_\mu} q_\mu^{n_\mu}\sum_{\bm{s}} C_{\bm{m},\bm{n}}^{(\bm{s})},\label{Ocl_HighDim}\\
    &O'_{sc}(\bm{q},\bm{p})=\sum_{\bm{m},\bm{n}}\sum_{\bm{s}} C_{\bm{m},\bm{n}}^{(\bm{s})}\left(Q^{(\bm{s}) }_{\bm{m},\bm{n}}\right)_{\rm sc}(\bm{p},\bm{q}),
  \end{align}
\end{subequations}
where
\begin{subequations}
  \begin{align}
    &\left(Q^{(\bm{s}) }_{\bm{m},\bm{n}}\right)_{\rm sc}(\bm{p},\bm{q})=-\prod_{\mu=1}^{f}p_\mu^{m_\mu} q_\mu^{n_\mu}\notag\\
    &\qquad\qquad+\prod_{\mu=1}^{f}\left[p_\mu^{m_\mu} q_\mu^{n_\mu}+\left(Q^{(s_\mu)}_{m_\mu,n_\mu}\right)_{\rm sc}(p_\mu,q_\mu)\right],\\
    &\left(Q^{(s_\mu)}_{m_\mu,n_\mu}\right)_{\rm sc}(p_\mu,q_\mu)\notag\\
    &\qquad =\sum_{\lambda=1}^{\min(m_\mu,n_\mu)} (i\hbar)^\lambda D_{m_\mu-\lambda,n_\mu-\lambda}^{(s_\mu)} p^{m_\mu-\lambda} q^{n_\mu-\lambda},
  \end{align}
\end{subequations}
and $D_{m_\mu-\lambda,n_\mu-\lambda}^{(s_\mu)}$ are coefficients 
 [see Eq.(\ref{D_Coefficient})].
 Clearly, $O'_{sc}(\bm{q},\bm{p})\rightarrow 0$ in the limit of $\hbar\rightarrow 0$.

 Making use of Eqs.(\ref{A_WeylO2A_Weylcl_HighDim}) 
 and (\ref{O_Weyl_Modify_HighDim})-(\ref{ww-Q-cl-g}), one gets that
\begin{equation}\label{Oqq'_2_Osc_HighDim}
  \bra{\bm{q}'}O(\hat{\bm{p}},\hat{\bm{q}})\ket{\bm{q}}=\frac{1}{(2\pi\hbar)^f}\int d\bm{p}\ e^{i \bm{p} 
  \cdot \bm{r}/\hbar}O_{\rm sc}(\bm{p},\bm{q}_0),
\end{equation}
 where $\bm{r}$ and $\bm{q}_0$ were defined in Eq.(\ref{q0-r}).
 Finally, substituting Eq.(\ref{Oqq'_2_Osc_HighDim}) into Eq.(\ref{Oii-1}), one finds that
\begin{equation}\label{OEi_Semiclassical_HighDim}
  O(E_i)=\int d\bm{p} d\bm{q}_0 \overline{W}_i(\bm{p},\bm{q}_0)O_{\rm sc}(\bm{p},\bm{q}_0).
\end{equation}
 Note that the result in Eq.(\ref{OEi_Semiclassical_HighDim}) is rigorous
 and is valid for an arbitrary quantum system that possesses a classical counterpart. 
 Approximating the averaged Wigner function by the energy surface as given in Eq.(\ref{eq-WFchaos})
 for quantum chaotic systems, one gets that
\begin{equation}\label{OEi_Semiclassical_HighDim-2}
  O(E_i) \simeq \frac 1{S({E_i})} \int d\bm{p} d\bm{q} \delta(H_{\rm cl}(\boldsymbol{p},\boldsymbol{q})-E_i)
  O_{\rm sc}(\bm{p},\bm{q}).
\end{equation}

\section{The offdiagonal function $f(e,e')$}\label{sect-semi-f(e,e')}
 
 In this section, we discuss a semiclassical approach to the function $f(e,e')$ 
 for offdiagonal elements in the ETH ansatz.
 The setup is given in Sec.\ref{section-f-setup}.
 The semiclassical study is given in Sec.\ref{sect-f-platform},
 which is based on Berry's conjecture on the action basis 
 discussed in Sec.\ref{Berry_Conjecture_Action},
 with the correlation functions neglected.

\subsection{Setup}\label{section-f-setup}

 Unlike discussions given in the previous section for the function $O(e)$, 
 which is valid for operators of both the considered system and its subsystems,
 in a study of the function $f(e,e')$, one usually considers an operator $O$ of a subsystem. 
 Indeed, the second term on the rhs of Eq.(\ref{ETH}) is not necessarily 
 valid for an operator of the total system.

 We use $S$ to indicate the considered quantum (conservative and chaotic) system,
 which possesses properties discussed in Sec.\ref{semi-EFs}. 
 For a conservative system, the chaoticity requires $f \ge 2$.
 The system $S$ may be divided into two subsystems indicated as
 $A$ and $B$, $S=A+B$.
 We use $O^A$ to indicate an arbitrary observable of $A$, with elements $O^A_{ij} = \la E_i|O^A|E_j\ra$.

 The configuration spaces of the classical counterparts of $A$ and $B$
 have dimensions denoted by $f_A$ and $f_B$, respectively,
 with coordinates indicated as  $({\bs q}_A, {\bs p}_A)$ and $({\bs q}_B, {\bs p}_B)$.
 Classical actions for subsystems are denoted by ${\bs I}_A$ and ${\bs I}_B$, respectively, 
 and the total action is written as ${\bs I}  = ({\bs I}_{A}, {\bs I}_{B})$.
 The corresponding angle variables are written as ${\bs \theta}  = ({\bs \theta}_{A}, {\bs \theta}_{B})$.

 The action operators $\hat{\bs I}_A$ and $\hat{\bs I}_B$ have discrete eigenvalues,
 which we indicate as ${\bs I}_{\bs a}$ and ${\bs I}_{\bs b}$, respectively, with discrete indices ${\bs a}$ and ${\bs b}$.
 The corresponding eigenstates are written as $|{\bs I}_{\bs a} \ra$ and $|{\bs I}_{\bs b}\ra$, respectively.
 Then, the total action operator $\hat{\bs I} $ has eigenstates $|{\bs I}_{\bs{ab}}\ra \equiv 
 |{\bs I}_{\bs a}\ra |{\bs I}_{\bs b}\ra$,
 with eigenvalues ${\bs I}_{\bs{ab}} = ({\bs I}_{\bs{a}}, {\bs I}_{\bs{b}})$.
 On the basis of $|{\bs I}_{\bs{ab}}\ra$, the eigenstates $|E_i\ra$ of the system $S$ are expanded as
 \begin{equation}\label{Ei-expan}
 |E_i\ra = \sum_{\bs{a}, \bs{b}} C^i_{\bs{ab}} |{\bs I}_{\bs{ab}}\ra
 \end{equation}
 with coefficients $C^i_{\bs{ab}}$.
 In consistency with the notation used in Sec.\ref{Berry_Conjecture_Action}, 
 one has $\bm{n} = (\bm{a}, \bm{b})$ and ${\bs I}_{\bs{n}} = {\bs I}_{\bs{ab}}$.

\subsection{A semiclassical approach to the function $f$}\label{sect-f-platform}

 We use $O_{ij}^{\rm off}$ to indicate offdiagonal elements of the operator $O^A$ on the total-energy basis,
\begin{align}\label{Oij-off}
 O_{ij}^{\rm off} = \la E_i |O^A|E_j\ra \quad \text{with $i \neq j$}.
\end{align}
 According to Eq.(\ref{eq-C9}), the EFs are written as
\color{black}
\begin{align}\label{Ciab-Pi-R}
 C^i_{\bs{ab}} \simeq  \hbar^{f/2} \sqrt{\Pi(E_i, {\bs I}_{\bs{ab}})} R^i_{\bs{ab}},
\end{align}
 where $R^i_{\bs{ab}}$  obeys  a Gaussian distribution {(zero mean and unit variance)}. 
 Making use of Eqs.(\ref{Ei-expan})-(\ref{Ciab-Pi-R}), one gets that
\begin{align}
 & O_{ij}^{\rm off} \simeq \hbar^f \sum_{\bs{a}, \bs{a}',{\bs b}}  
 \sqrt{\Pi(E_i, {\bs I}_{\bs{ab}}) \Pi(e_{j(\ne i)}, {\bs I}_{\bs{a}' \bs{b}})} 
 R^{i*}_{\bs{ab}}R^{j}_{\bs{a}' \bs{b}} O^{A}_{\bs{a} \bs{a}'},
 \label{O3}
\end{align}
 where $ O^{A}_{\bs{a} \bs{a}'} = \la {\bs I}_{\bs{a}}|  O^{A} |{\bs I}_{{\bs a}'}\ra$.
 The average value of ${\abs{O_{ij}^{\rm off}}^2}$, namely $\overline{\abs{O_{ij}^{\rm off}}^2}$,
 gives the function $f(e,e')$.

 As discussed previously,
 components within EFs usually possess nonvanishing correlations, 
 implying nonvanishing $\ov{R^i_{\bs{ab}}R^i_{\bs{a}' \bs{b}'}}$.
 Furthermore, this type of correlation, in fact, 
 implies some nonvanishing correlations between different EFs, 
 i.e., nonvanishing $\ov{R^i_{\bs{ab}}R^j_{\bs{a}' \bs{b}'}}$ of $i\ne j$,
 even though a random global phase may be assigned to each EF. 
 Indeed, there are totally $d_\HH$ such global phases, where $d_\HH$ indicates the 
 dimension of the Hilbert space, while, the number of $\ov{R^i_{\bs{ab}}R^j_{\bs{a}' \bs{b}'}}$
 scales as $d_\HH^2$.

 Clearly, derivation of a useful expression for $\ov{R^i_{\bs{ab}}R^j_{\bs{a}' \bs{b}'}}$
 should be even harder than that for $\ov{R^i_{\bs{ab}}R^i_{\bs{a}' \bs{b}'}}$.
 Below, we are to derive a semiclassical expression for the quantity $\overline{\abs{O_{ij}^{\rm off}}^2}$, 
 by neglecting the autocorrelation functions of $\ov{R^i_{\bs{ab}}R^j_{\bs{a}' \bs{b}'}}$.
 \footnote{In the next section, we are to study numerically whether such an expression 
 could supply  information that are useful to some extent.}

 Assuming that $R^{i}_{\bs{ab}}$ are independent random numbers  
 with $\overline{\abs{R^i_{\bm{ab}}}^2}=1$, after averaging over $i$ and $ j$, 
 from Eq.(\ref{O3}), one gets that
\begin{equation}\label{Ooff^2}
  \overline{\abs{O_{ij}^{\rm off}}^2}
   \simeq \hbar^{2f}\sum_{\bm{a},\bm{a}',\bm{b}}\Pi\left(E_i,\bm{I_{ab}}\right)\Pi\left(E_j,
  \bm{I_{a'b}}\right)\abs{O^A_{\bm{aa}'}}^2.
\end{equation} 
 Noting Eq.(\ref{Pi-E-I}), this gives that
 \begin{equation}\label{O3^2-app} 
  \begin{aligned}
    &\overline{\abs{O_{ij}^{\rm off}}^2}
     \simeq \sum_{\bs{a}, \bs{a}', \bs{b}} \frac{\hbar^{2f}  \abs{O^{A}_{\bs{a} \bs{a}'}}^2 }{S(E_i) S(E_j)}\\
    &\times\int d\boldsymbol{p}d\boldsymbol{q}\delta(E_i-H_{\rm cl}(\boldsymbol{p}, \boldsymbol{q}))
          \delta^f({\bs I}_{\bs{ab}}-{\bs I}_{\rm cl}(\boldsymbol{p},\boldsymbol{q}))\\
    &\times\int d\boldsymbol{p}'d\boldsymbol{q}'\delta(E_j-H_{\rm cl}(\boldsymbol{p}',
    \boldsymbol{q}'))
     \delta^f({\bs I}_{\bs{a}' \bs{b}}-{\bs I}_{\rm cl}(\boldsymbol{p}',\boldsymbol{q}')).
  \end{aligned}
\end{equation}

 In the semiclassical limit of $\hbar \to 0$,
 one may change the summation over ${\bs a}$ and ${\bs b}$
 to an integration over the classical action ${\bs I} $,
 i.e., $\sum_{\bs{a}, \bs{b}} \hbar^{f} \to \int d{\bs I} $.
 Then, Eq.(\ref{O3^2-app}) becomes
\begin{equation}\label{OA^2-fin1}
  \begin{aligned}
    &\overline{\abs{O_{ij}^{\rm off}}^2}
     \simeq  \hbar^{f_B}\int  \frac{d\widetilde{\bm{I}}^A d\widetilde{\bm{I}}'^A d\widetilde{\bm{I}}^B }{S(E_i) S(E_j)}  
    \left| O^{A}_{\rm qm}(\ww{\bs{I}}^A, \ww{\bs{I}}'^A) \right|^2
   \\ &  \times  \int d\boldsymbol{p}d\boldsymbol{q}\delta(E_i-H_{\rm cl}(\boldsymbol{p},\boldsymbol{q}))
   \delta^{f_A}\left(\widetilde{\bm{I}}^A -{\bs I}^A_{\rm cl}(\boldsymbol{p},\boldsymbol{q})\right)\\
   &\qquad\qquad\qquad\qquad\times\delta^{f_B}\left(\widetilde{\bm{I}}^B -{\bs I}^B_{\rm cl}(\bm{p},\bm{q})\right)
    \\ &  \times \int d\boldsymbol{p}'d\boldsymbol{q}'\delta(E_j-H_{\rm cl}(\boldsymbol{p}',\boldsymbol{q}'))
   \delta^f\left(\widetilde{\bm{I}}'^{A} -{\bs I}^A_{\rm cl}(\boldsymbol{p}',\boldsymbol{q}')\right)\\
   &\qquad\qquad\qquad\qquad\times\delta^{f_B}\left(\widetilde{\bm{I}}^B -{\bs I}^B_{\rm cl}(\bm{p}',\bm{q}')\right),
  \end{aligned}
\end{equation}
 where $O^{A}_{\rm qm}(\ww{\bs{I}}_A, \ww{\bs{I}}'_A)$ is obtained by writing $O^A_{\bs{a} \bs{a}'}$
 as a function of $({\bs I}_{\bs a}, {\bs I}_{{\bs a}'})$ and then replacing $({\bs I}_{\bs a}, {\bs I}_{{\bs a}'})$ by
 $(\ww{\bs I}_{A}, \ww{\bs I}'_{A})$.
 Carrying out the integrations over $\ww{\bs I}^A$, $\ww{\bs I}'^A$ and $\widetilde{\bm{I}}^B$, one gets that
\begin{equation}\label{OA^2-fin} 
  \begin{aligned}
     &\overline{\abs{O_{ij}^{\rm off}}^2}
     \simeq  \hbar^{f_B}\int \frac{ d\boldsymbol{p}d\boldsymbol{q}  d\boldsymbol{p}'d\boldsymbol{q}'}{S(E_i) S(E_j)}
    \abs{O^{A}_{\rm qm}\left({\bs I}_{A,\rm cl},  {\bs I}'_{A,\rm cl}\right)}^2
    \\  & \times
    \delta^{f_B}\left(\bm{I}_{B,\rm cl}-\bm{I}'_{B,\rm cl}\right)
    \delta(E_i-H_{\rm cl}(\boldsymbol{p}, \boldsymbol{q}))
    \delta(E_j-H_{\rm cl}(\boldsymbol{p}', \boldsymbol{q}')).
  \end{aligned}
\end{equation}
 where $({\bs I}_{A,{\rm cl}},  {\bs I}_{B,{\rm cl}})$
 and $ ({\bs I}'_{A,{\rm cl}}, {\bs I}'_{B,{\rm cl}})$ are functions of
 $(\boldsymbol{p}, \boldsymbol{q})$ and $(\boldsymbol{p}',\boldsymbol{q}')$, respectively.

 Then, under a canonical transformation of $({\bs p}, {\bs q}, {\bs p}', {\bs q}')
 \to ({\bs \theta}, {\bs I}, {\bs \theta}', {\bs I}')$, Eq.(\ref{OA^2-fin}) gives that
\begin{equation}
  \begin{aligned}
    & \overline{\abs{O_{ij}^{\rm off}}^2}
     \simeq  \hbar^{f_B}\int \frac{ d\boldsymbol{I}d\boldsymbol{\theta}  
     d\boldsymbol{I}'d\boldsymbol{\theta}'}{S(E_i) S(E_j)}
   \left|O^{A}_{\rm qm} ({\bs I}_{A},  {\bs I}'_{A}) \right|^2 \\
   &\times\delta^{f_B}({\bs I}_{B}-  {\bs I}'_{B})
   \delta(E_i-H_{\rm cl}(\bs{I}, \bs{\theta})) \delta(E_j-H_{\rm cl}(\bs{I}', \bs{\theta}')).
  \end{aligned}
\end{equation}
 Finally, carrying out the integration over ${\bs I}_B$, one gets the following semiclassical result for 
 the function $f(e,e')$, indicated as $f_{\rm sc}(e,e')$,
\begin{align}
  & f_{\rm sc}^2(E_i,E_j) =  \frac{\hbar^{f_B}}{S(E_i) S(E_j)} \int d\boldsymbol{I}_A d\boldsymbol{\theta}
  F_{ij}(\bs{I}_A, \bs{\theta}),
\label{g^2-action}
\end{align}
 where
\begin{align}\label{}
 \notag   F_{ij}( \bs{I}_A, \bs{\theta})  & =  \int d\boldsymbol{I}' d\boldsymbol{\theta}'
 \delta(E_j-H_{\rm cl}(\bs{I}', \bs{\theta}'))
 \\ & \times \delta(E_i-H_{\rm cl}(\bs{I}_A, \bs{I}'_B, \bs{\theta}))
 |O^{A}_{\rm qm}({\bs I}_{A},  {\bs I}'_{A})|^2.
 \label{g-def}
\end{align}

\section{Numerical simulations}\label{sect-numerical}

 In this section, we discuss numerical simulations that have been carried out for the following three purposes.
 (i) To test the semiclassical prediction given in Sec.\ref{sect-semi- O(e)} for the diagonal function $O(e)$;
 (ii) to show that the offdiagonal function $f(e,e')$, which is computed in a chaotic system
 with two degrees of freedom, demonstrates qualitative features 
 of those previously observed in many-body systems;
 and (iii) to study the extent of validity of the semiclassical result given in Sec.\ref{sect-semi-f(e,e')}.

 Specifically, preliminary discussions are given in Sec.\ref{sect-num-prelimi}.
 The model to be employed is discussed in Sec.\ref{sect-LMG-model}.
 And, numerical simulations are discussed in Sec.\ref{sect-numeric-LMG}.

\begin{figure}[!t]
  \centering
  \includegraphics[width=1\linewidth]{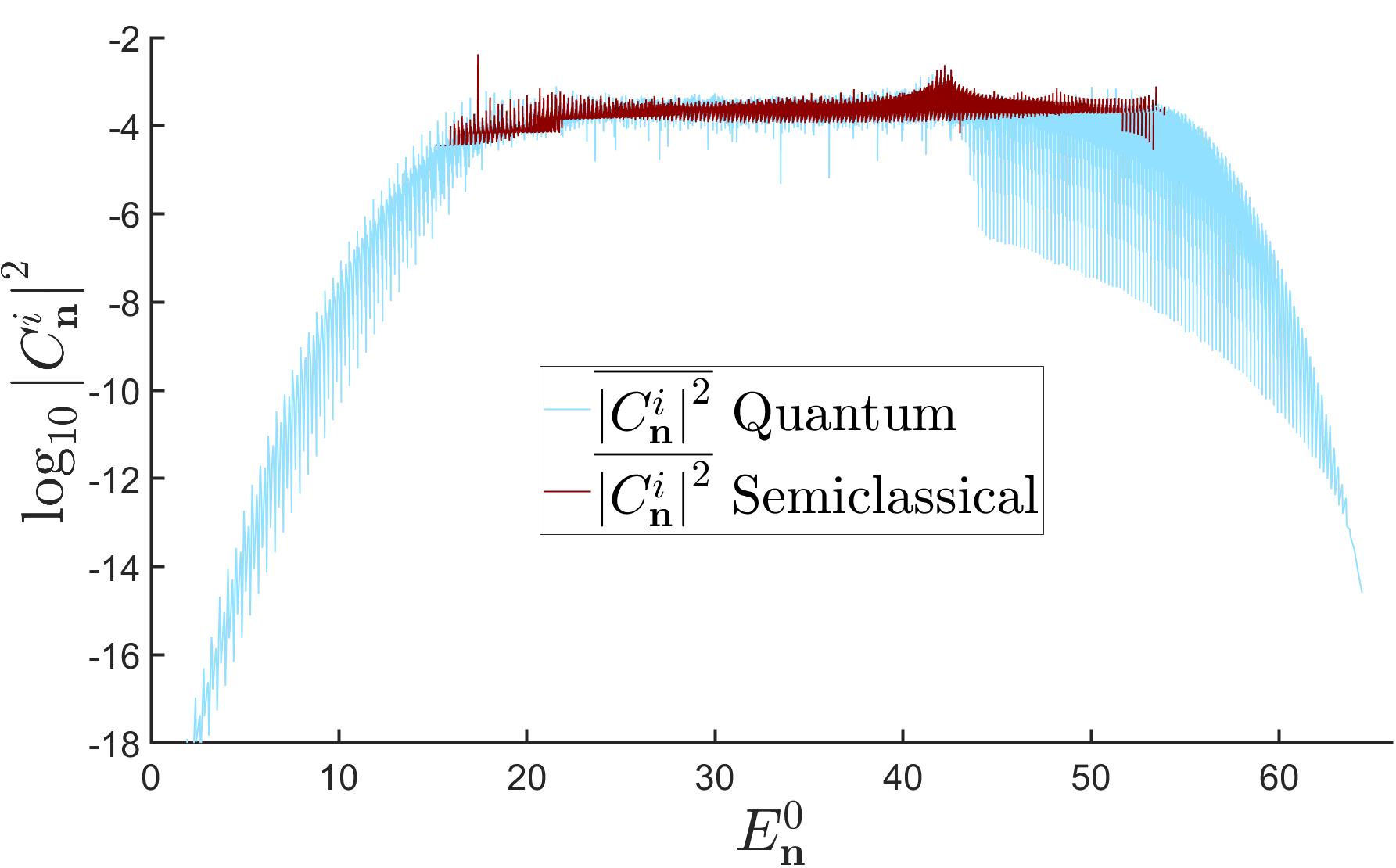}
  \caption{ $\ov {|C^{i}_{\bs{n}}|^{2}}$, average shape of EFs in the middle energy region of the LMG model 
 as a chaotic system, in the logarithm scale. 
 Average was taken over about 50 energy levels $E_i$
  in the middle energy region.
  Parameters are  $\epsilon'_{1}=44.00,  \epsilon'_{2}= 64.40,
 \mu'_{1}= 18.56, \mu'_{2} = 27.40, \mu'_{3}= 25.28$, $\mu'_{4} = 7.024$,
 $\lambda=2$, and $\Omega=100$. 
 The semiclassical predictions were computed by Eq.(\ref{eq-EFshapecl}),
 which show agreement with the exact values in an average sense in the 
 classically and energetically allowed region. 
}
  \label{EFs_Shape}
\end{figure}

\subsection{Preliminary discussions}\label{sect-num-prelimi}

 The ETH ansatz was proposed for understanding thermalization of many-body systems.
 However, the many-body feature should not be crucial for 
 all properties of the two functions of $O(e)$ and $f(e,e')$.
 This suggests that, for the purpose of understanding some aspects of them,
 models with a few degrees of freedom may be useful.

 To be specific, firstly, one observes that the dimension $f$ does not explicitly 
 appear in the semiclassical expression for the function $O(e)$,
 which is given in Eq.(\ref{OEi_Semiclassical_HighDim})
 (see also the special case in Eq.(\ref{O(E)-sc}) for the limit of $\hbar \to 0$).
 This suggests that one may employ a quantum chaotic model whose degrees of freedom is small
 ($f\ge 2$), 
 for the purpose of numerically testing the semiclassical prediction.

 Secondly, as is known from previous numerical simulations (see discussions given in Sec.\ref{sect-introduction}),
 quantitative behaviors of the function $f(e,e')$ show $f$-dependent features,
 particularly, height and width of its platform in the region of $e$ close to $e'$
 \cite{Rigol_2020,Vidmar_2019,Rigol_2019,Rigol_2021}.
 One question of relevance is whether the function $f(e,e')$, 
 which is computed in a quantum chaotic system with a few degrees of freedom, 
 may exhibit behaviors that are qualitatively similar to those observed in many-body systems. 
 If this could be possible, as to be shown below in Sec.\ref{sect-numeric-LMG}, 
 then, for the purpose of understanding qualitatively behaviors of $f(e,e')$,
 one may study systems of the former type, which is much easier than the latter. 

\begin{figure}[!t]
  \centering
  \includegraphics[width=1\linewidth]{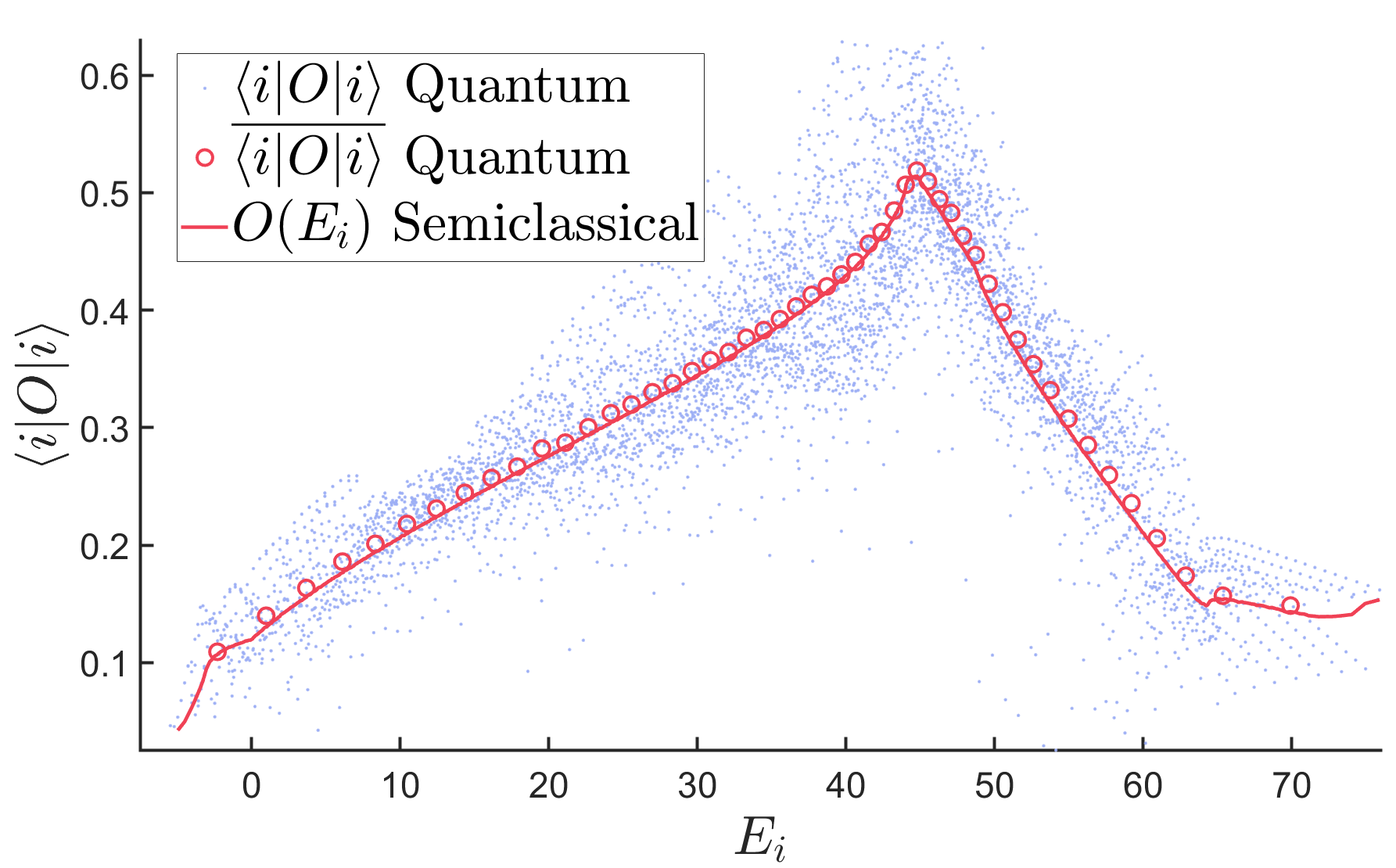}
  \caption{ Values of the diagonal elements $\la i|O|i\ra $ for $O=K_{11}/\Omega$
  (blue dots), their locally-averaged values obtained by averaging over about 100 neighboring levels (red open circles), 
 and the semiclassical prediction given in Eq.(\ref{OEi_Semiclassical_HighDim-2}) (red line).}
  \label{f_Compare}
\end{figure}

\subsection{The LMG Model}\label{sect-LMG-model}

The model we employ is a three-orbital LMG model \cite{Lipkin_1965}.
 This model is composed of $\Omega$ particles, occupying three energy levels labeled by $r=0,1,2$, each with
 $\Omega$-degeneracy.
 Here, we are interested in the collective motion of this model, for which the dimension of the
 Hilbert space is $\frac 12 (\Omega+1)(\Omega +2)$  \cite{WIC98}.
 We use $\epsilon_{r}$ to denote the energy of the $r$-th level
 and, for brevity, set $\epsilon_{0}=0$.

 The Hamiltonian of the model is written as
\begin{equation}\label{H_LMG}
H=H_0+\lambda V,
\end{equation}
 where $H_0$ indicates the Hamiltonian of an integrable system and $V$ is a perturbation.
 Specifically,
\begin{gather}
 H_{0}=\epsilon_{1}K_{11}+\epsilon_{2}K_{22},  \\
V=\sum_{t=1}^{4}\mu_{t}V^{(t)},
\end{gather}
where
\begin{equation}
  \begin{aligned}
    V^{(1)}=K_{10}K_{10}+K_{01}K_{01},\ V^{(2)}=K_{20}K_{20}+K_{02}K_{02},\\
 V^{(3)}=K_{21}K_{20}+K_{02}K_{12},\ V^{(4)}=K_{12}K_{10}+K_{01}K_{21}.
  \end{aligned}
\end{equation}
Here, the operators $K_{rs}$ are defined by
\be
K_{rs}=\sum_{\gamma=1}^{\Omega}a_{r\gamma}^{\dagger}a_{s\gamma},\quad r,s=0,1,2,
\ee
where $a^{\dagger}_{r\gamma}$ and $a_{r\gamma}$ are fermionic creation and annihilation
operators obeying the usual anti-commutation relations.

 For symmetric states, the operators $K_{rs}$ can be written
 in terms of bosonic creation and annihilation operators $b^\dagger _r$ and $b_r$ \cite{Gong-ou_Xu_1995},
\be
K_{rs}=b_{r}^{\dagger}b_{s},\quad K_{r0}=K_{0r}^{\dagger}=b_{r}^{\dagger}
\sqrt{\Omega-b_{1}^{\dagger}b_{1}-b_{2}^{\dagger}b_{2}},
\ee
for $r,s=1,2$.
 Under the transformation,
\begin{equation}
b_{r}^{\dagger}=\sqrt{\frac{\Omega}{2}}(\hat{q}_{r}-i\hat{p}_{r}),\ \ \ b_{r}=\sqrt{\frac{\Omega}{2}}(\hat{q}_{r}+i\hat{p}_{r}),
\end{equation}
for $r=1,2$, it is easy to verify that $q_{r}$ and $p_{s}$ obey the following commutation relation,
\begin{equation}
[q_{r},p_{s}]=\frac{i}{\Omega}\delta_{rs}.
\end{equation}
 Hence, $1/\Omega$ plays the role of an effective Planck constant,
\be
\hbar_{\rm eff}=\frac{1}{\Omega}.
\ee

\begin{figure}[!t]
  \centering
  \includegraphics[width=1\linewidth]{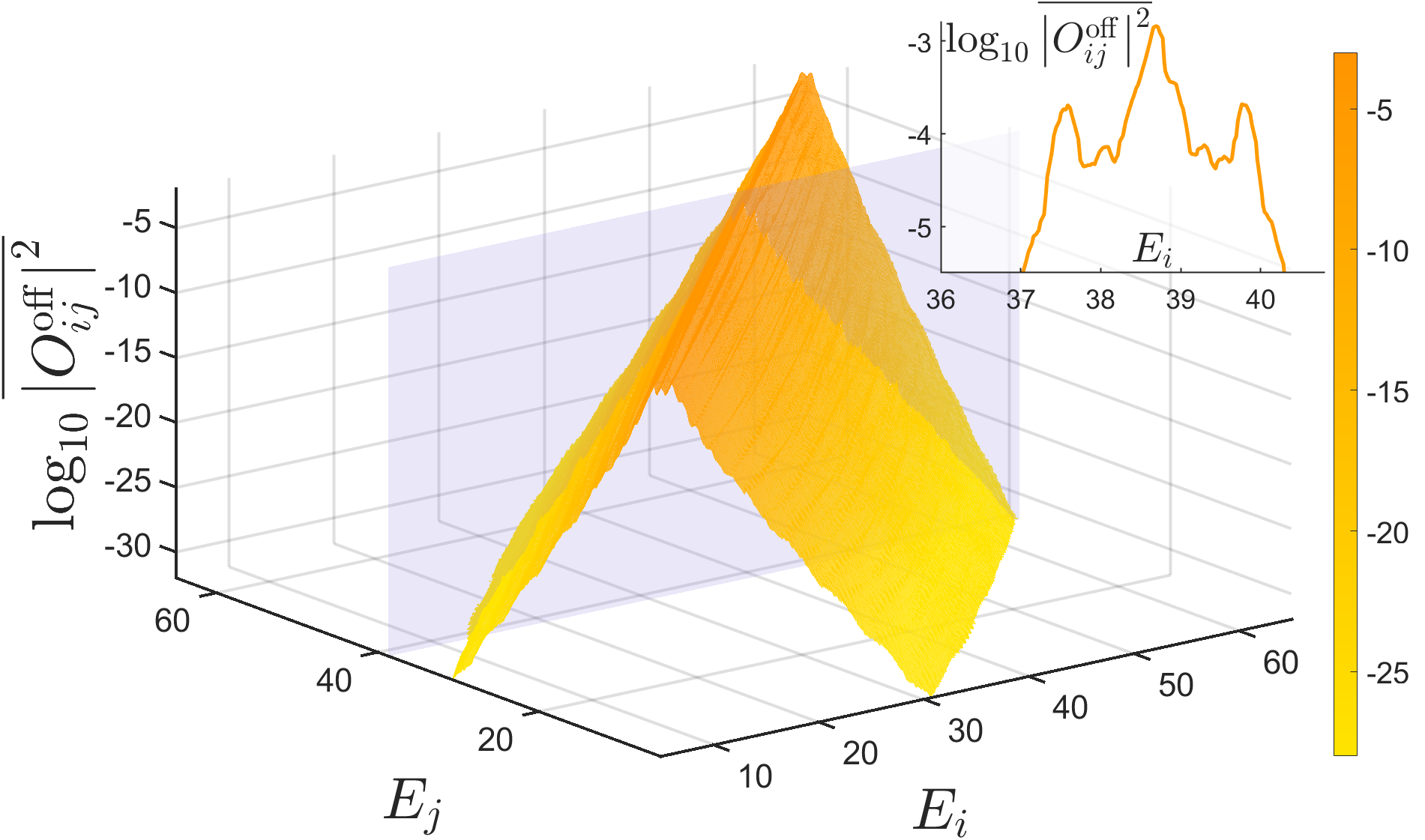}
  \caption{Variation of $\overline{\abs{O_{ij}^{\rm off}}^2}$ for offdiagonal elements versus $E_i$ 
  and $E_j$, with average taken over narrow energy shells each containing 15 levels round $E_i$ and $E_j$.
  The inset shows the top region of a section taken at $E_j=38.69$ (indicated by the purple plane).
  }
  \label{g2_LMG_Quantum}
\end{figure}

 It is straightforward to find that the classical counterpart of the model,
 which possesses a two-dimensional configuration space ($f=2$), 
 has the following Hamiltonian \cite{WIC98,Meredith88},
\be
H(\boldsymbol{p},\boldsymbol{q})=H_0(\boldsymbol{p},\boldsymbol{q})
 + \lambda V(\boldsymbol{p},\boldsymbol{q}),
\ee
where
\begin{gather}
H_{0}(\boldsymbol{p},\boldsymbol{q})=\frac{\epsilon'_{1}}{2}(p_{1}^{2}+q_{1}^{2})+\frac{\epsilon'_{2}}{2}(p_{2}^{2}+q_{2}^{2}), \nonumber \\
V(\boldsymbol{p},\boldsymbol{q})=\mu'_{1}(q_{1}^{2}-p_{1}^{2})(1-G/2)+\mu'_{2}(q_{2}^{2}-p_{2}^{2})(1-G/2) \nonumber \\
+\frac{\mu'_{3}}{\sqrt{2}}[(q_{2}^{2}-p_{2}^{2})q_{1}-2q_{2}p_{1}p_{2}]\sqrt{1-G/2} \nonumber \\
+\frac{\mu'_{4}}{\sqrt{2}}[(q_{1}^{2}-p_{1}^{2})q_{2}-2q_{1}p_{1}p_{2}]\sqrt{1-G/2},
\end{gather}
with $G=q_{1}^{2}+p_{1}^{2}+q_{2}^{2}+p_{2}^{2}\le2$.
 Here,  the classical parameters are given by
 $\epsilon'_{1}=\epsilon_{1} \Omega, \epsilon'_{2}=\epsilon_{2} \Omega,
 \mu'_{1} = \mu_{1} \Omega^2, \mu'_{2} = \mu_{2} \Omega^2, \mu'_{3}= \mu_{3} \Omega^2$,
and $\mu_{4}'=\mu_{4} \Omega^2$.

 In numerical simulations, fixed classical parameters were used, 
 which are $\epsilon'_{1}=44.00,  \epsilon'_{2}= 64.40,
 \mu'_{1}= 18.56, \mu'_{2} = 27.40, \mu'_{3}= 25.28$, $\mu'_{4} = 7.024$,
$\lambda=2$, and $\Omega=100$.
 Under these parameters, different values of 
 the particle number $\Omega$ correspond to a same classical counterpart, which is chaotic.
 Properties of the LMG model in the quantum chaotic region have been studied 
 well in previous works (e.g., see Refs.\cite{Meredith88,WIC98,pre18-EF-BC}).
 Here, we are to discuss only properties of relevance to the understanding of ETH.

 Numerically, we have checked 
 the semiclassical prediction given in Eqs.(\ref{eq-EFshapecl})-(\ref{eq-ov1}) 
 for the average shape of EFs in  classically and energetically allowed region
 (Fig.\ref{EFs_Shape} as an example).
 And, the distribution of rescaled EF components $R^i_{\bm{n}}$ [see Eq.(\ref{eq-C9})]
 were found quite close to the Gaussian distribution, as reported previously \cite{pre18-EF-BC}.

\subsection{Numerical simulations for the two functions}\label{sect-numeric-LMG}

 Let us first discuss numerical test for the semiclassical prediction of the diagonal 
 function $O(e)$ given in Eq.(\ref{OEi_Semiclassical_HighDim-2}).
 For this purpose, we have studied an operator $O$ as $O= K_{11}/\Omega$,
 for which $O_{\rm sc}(\bm{p},\bm{q})=\frac{1}{2}(p_1^2+q_1^2)-\frac{1}{2}\hbar_{\rm eff}$
 given by Eq.(\ref{Osc_Definition}).
 In Fig.\ref{f_Compare}, it is seen that 
 the semiclassical prediction of $O(e)$ given by Eq.(\ref{OEi_Semiclassical_HighDim-2}) works well (red line), 
 being close to the locally averaged values of $\la i|O|i\ra $ (red open circles) 
 computed numerically in the quantum model.

 Next, we study whether the averaged offdiagonal elements $\ov{\abs{O_{ij}^{\rm off}}^2}$ in the LMG model
 may show qualitatively similar behaviors as the offdiagonal function $f(e,e')$ observed in many-body
 models satisfying ETH. 
 In the LMG model, the degree of freedom of $q_1$ is taken as the subsystem $A$ and that of $q_2$ as $B$. 
 We still study the operator of $O = O^A = K_{11}/ \Omega$.

 Variation of $\overline{\abs{O_{ij}^{\rm off}}^2}$, as a function of $E_i$ and $E_j$, is plotted 
 in Fig.\ref{g2_LMG_Quantum} in the logarithm scale, 
 where the average was taken over narrow energy shells around $E_i$ and $E_j$.
 It shows a quite good exponential decay with increasing $|E_i-E_j|$, 
 which is similar to what is known in many-body models discussed in Sec.\ref{sect-introduction}. 
 To see more clearly the behavior   of $\overline{\abs{O_{ij}^{\rm off}}^2}$ at $E_i$ close to $E_j$, 
 the top region of the section at $E_j=38.69$ is plotted in the inset,
 where loosely speaking  a platform-type shape is seen.
 Compared with platforms observed in many-body models, 
 this shape is narrow and shows large fluctuations, which may be due to the smallness of $f$.

 The above numerical results show that, although the LMG model possesses only two degrees of freedom, 
 the quantity of $\overline{\abs{O_{ij}^{\rm off}}^2}$ in this model behaves 
 in a way qualitatively similar to that found in many-body models satisfying ETH. 
 In other words, quantum chaotic models with a few degrees of freedom 
 may supply useful information in the study of the function $f(e,e')$ in the ETH ansatz.

\begin{figure}[!t]
  \centering
  \includegraphics[width=1\linewidth]{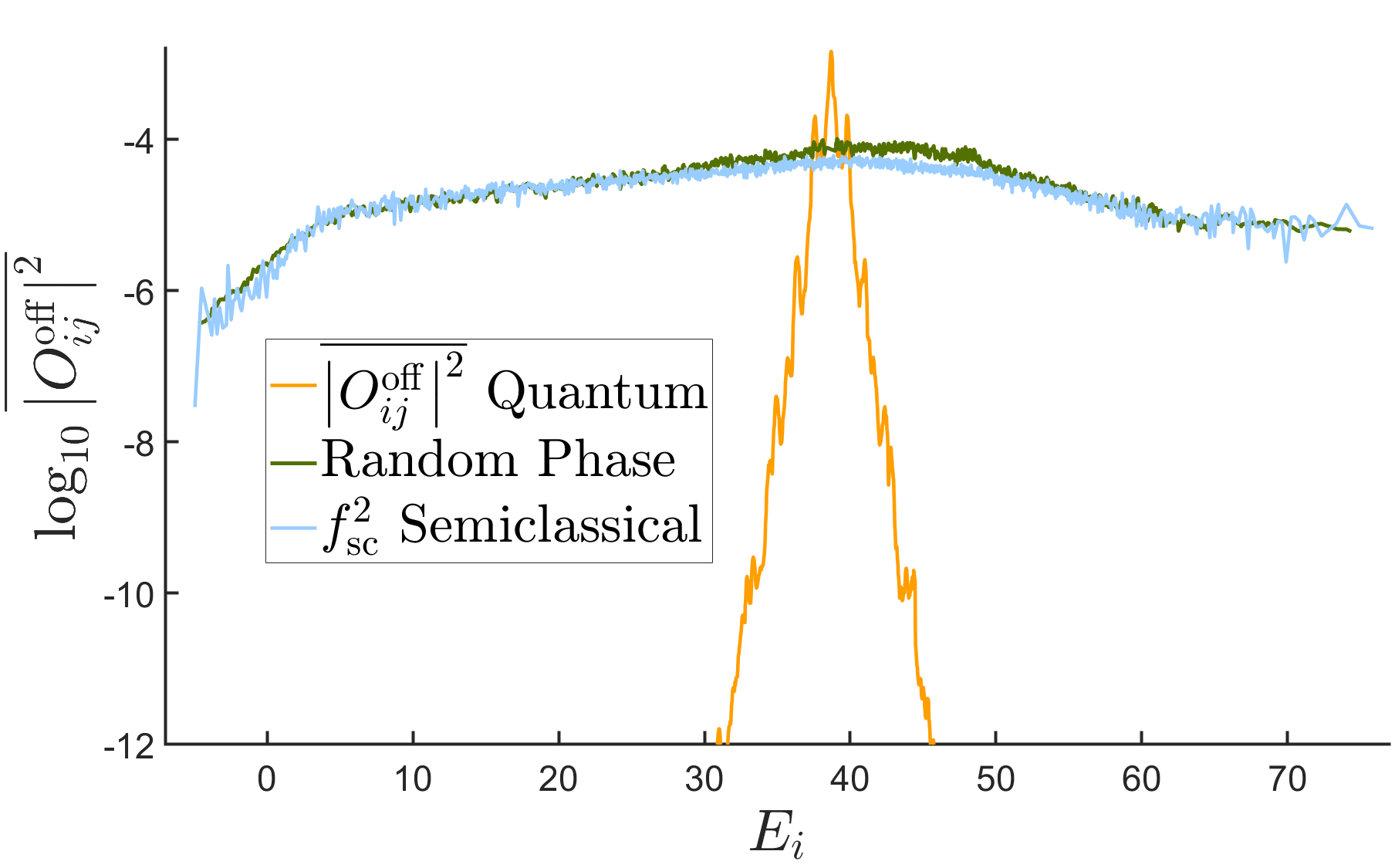}
  \caption{A section of Fig.\ref{g2_LMG_Quantum} taken at $E_j=38.69$.
 The blue line indicates the semiclassical prediction of Eq.(\ref{g^2-action})
 and the green line indicates $\overline{\abs{\bra{i_R}O\ket{j_R}}^2}$ for the phase-randomized states
 $\ket{i_R}$ in Eq.(\ref{Randomized_State}).
  }
  \label{g2_Compare1}
\end{figure}

 Then, we discuss whether the semiclassical result given in Eq.(\ref{g^2-action}),
 which was obtained under the assumption of negligible correlation functions of the EFs on 
 the action basis, 
 may be of any relevance for understanding properties of the function $f(e,e')$. 
 For this, we have plotted the whole section for the inset of Fig.\ref{g2_LMG_Quantum},
 as shown in Fig.\ref{g2_Compare1}, 
 where the blue line gives the semiclassical result of Eq.(\ref{g^2-action}). 
 It is seen that the prediction of Eq.(\ref{g^2-action}),
 which was derived without taking into account the correlation functions,
 decays much slower than the exact value of $\overline{\abs{O_{ij}^{\rm off}}^2}$.
 In other words, correlation functions of the EFs on the action basis should be  nonnegligible
 and give significant contributions to $O_{ij}^{\rm off}$ in Eq.(\ref{O3}),
 such that $\overline{\abs{O_{ij}^{\rm off}}^2}$ may exhibit a fast exponential decay.

 In order to check the above analysis in the relevance of the correlation functions, 
 we have studied impact of randomized phases of the EFs. 
 More exactly, instead of the eigenstates $\ket{i}=\sum_{\bm{n}} C^i_{\bm{n}}\ket{\bm{n}}$,
 whose coefficients $C^i_{\bm{n}}$ are real in the studied model,
 we have studied the following states, denoted by $\ket{{i}_R}$, 
\begin{equation}\label{Randomized_State}
 \ket{{i}_R} =\sum_{\bm{n}} (-1)^\eta \abs{C^i_{\bm{n}} }  \ket{\bm{n}}, 
\end{equation}
 where $\eta=0$ or $1$ chosen randomly with the variation of $i$ and $\bm{n}$.
 The corresponding values of $\overline{\abs{O_{ij}^{\rm off}}^2}$ for the phase-randomized states, 
 which were computed by replacing $|E_{i(j)}\ra$ in Eq.(\ref{Oij-off}) by $\ket{{i(j)}_R}$,
 are also plotted in Fig.\ref{g2_Compare1} (green line).
 As expected, they are close to the semiclassical results (blue line) computed by Eq.(\ref{g^2-action}).

\begin{figure}[!t]
  \centering
  \includegraphics[width=1\linewidth]{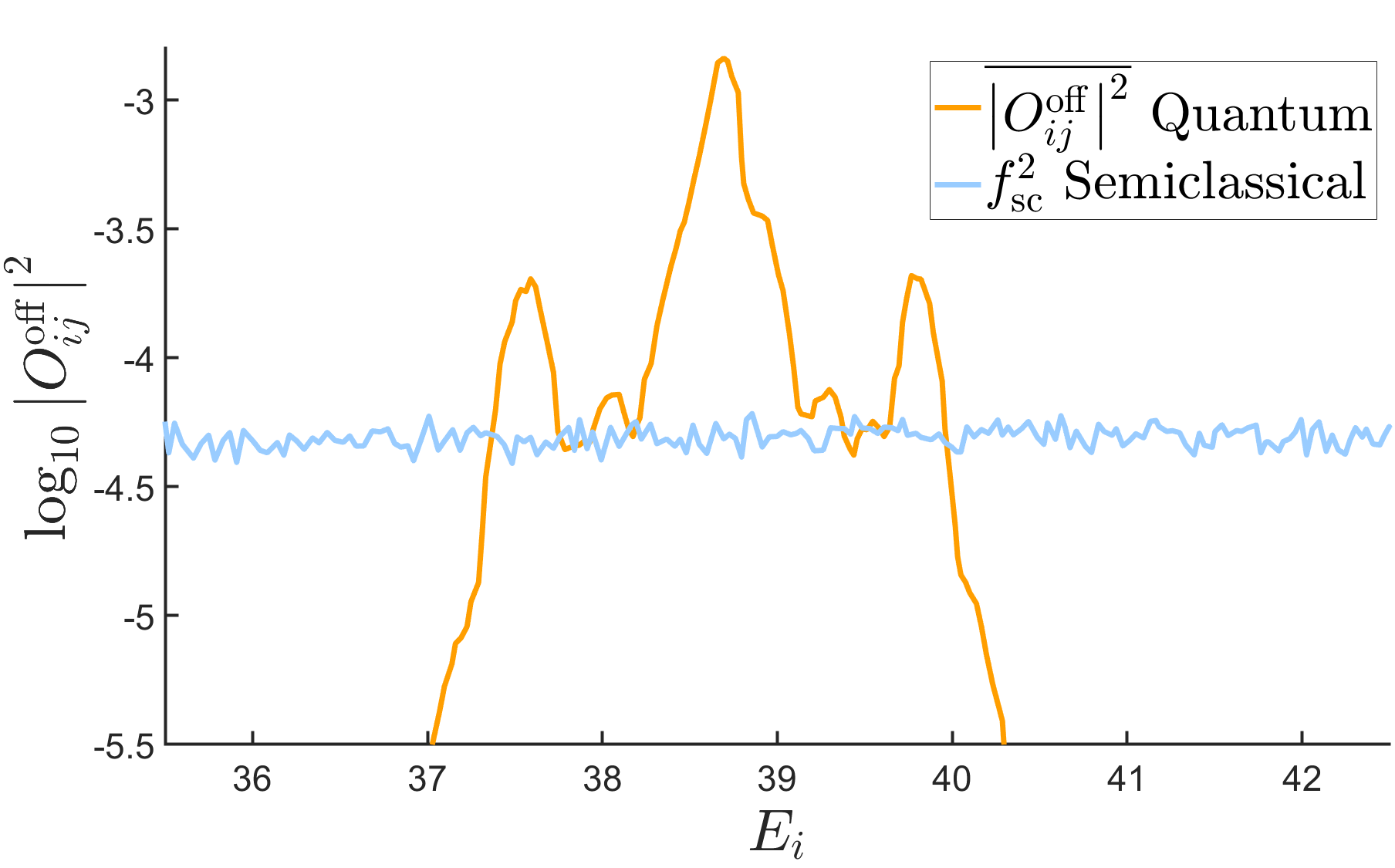}
  \caption{Comparison of the semiclassical prediction of Eq.(\ref{g^2-action}) and the exact values,
 as taken from the central region of Fig.\ref{g2_Compare1}.
  }
  \label{g2_Compare3}
\end{figure}

 Although the semiclassical prediction of Eq.(\ref{g^2-action}) deviates largely
 from the exact quantum values in the far-tail region with large $|E_i-E_j|$,
 as seen in Fig.\ref{g2_Compare1},  the deviation is not so large in the central region 
 with $E_i \approx E_j$. 
 To see the details, we have enlarged the central region, which contains the 
 above-discussed platform-type shape.
 In Fig.\ref{g2_Compare3},
 it is seen that Eq.(\ref{g^2-action}) works approximately in the central region,
 in the sense that it shares the same order of magnitude as the platform-type 
 shape of the function $f(e,e')$ in the LMG model. 
 In other words, the semiclassical prediction of Eq.(\ref{g^2-action}) may be useful 
 in an estimation for the scale of $f(e,e')$ in the platform region.

\section{Conclusions and discussions}\label{sect-conclusion}

\subsection{Summary}

 In this paper, three main results have been obtained. 
 The first one is a semiclassical expression, which is for the diagonal function
 $O(e)$ in the ETH ansatz.
 This expression is derived by a method, which makes use of the concept of Weyl-ordered operator
 and is much simpler than those used previously in the derivation of a 
 semiclassical expression valid in the limit of $\hbar \to 0$. 
 Even further, this method may be used to rigorously derive
 higher-order contributions of $\hbar$  from the operator aspect.

 The second result is based on numerical simulations
 that have been performed in the LMG model.
 That is, the function $f(e,e')$ for offdiagonal elements, 
 which is computed in a quantum chaotic system with a few degrees of freedom, 
 may show behaviors that are qualitatively similar to those observed in many-body systems satisfying 
 the ETH ansatz. 
 This suggests that few-body quantum chaotic systems may supply useful information
 in the study of ETH for many-body systems.

 The third result is a semiclassical expression for the offdiagonal function $f(e,e')$ in the ETH ansatz,
 which is derived under the assumption of negligible correlation functions of chaotic EFs on an action basis.
 Numerical simulations in the LMG model show that this expression may be useful 
 for estimating the height of the central ``platform'' of the function $f(e,e')$ 
 in the regime of $e \simeq e'$.

\subsection{Some discussions}\label{SubSec_OffD_Application}

 The above discussed numerical support to the third result suggests that 
 the semiclassical expression obtained for $f(e,e')$
 may useful for showing a $\rho_{\rm dos}^{-1/2}$ scaling behavior of $f(e,e')$ (in the platform region), 
 as one of the main contents of the ETH ansatz.
 However, a rigorous derivation is still difficult.
 Below, we give some qualitative arguments.

 For this purpose, one needs to give an estimate  to the integration on the rhs of Eq.(\ref{g-def}).
 In particular, a comparison is useful with the following expression of $S({E_j})$, which is obtained 
by a canonical transformation for the rhs of Eq.(\ref{SE}), 
\be \label{SE-action}
 S({E_j})=\int d\boldsymbol{I}' d\boldsymbol{\theta}'
 \delta(E_j-H_{\rm cl}(\boldsymbol{I}',\boldsymbol{\theta}')).
\ee
One notes that the integrand in Eq.(\ref{g-def}) does not vanish, only if the energy surface 
of $H_{\rm cl}(\bs{I}_A', \bs{I}_B', \bs{\theta}') = E_j$ coincides with that of 
$H_{\rm cl}(\bs{I}_A, \bs{I}'_B, \bs{\theta}) = E_i$, that is, if
\begin{align}\label{H-H-Eij}
 H_{\rm cl}(\bs{I}_A, \bs{I}'_B, \bs{\theta}) - H_{\rm cl}(\bs{I}_A', \bs{I}_B', \bs{\theta}') = E_i -E_j.
\end{align}
 Clearly, Eq.(\ref{H-H-Eij}) does not hold for an arbitrary point of $(\bs{I}', \bs{\theta}')$,
 with $\bs{I}' = (\bs{I}'_A, \bs{I}'_B)$, and a given point of $(\bs{I}_A, \bs{I}'_B, \bs{\theta})$.
 However, in the case $E_i \approx E_j$, for each given point of $(\bs{I}', \bs{\theta}')$,
usually there should exist coordinates $(\bs{I}_A, \bs{\theta})$ in the neighborhood of 
$(\bs{I}'_A, \bs{\theta}')$, for which Eq.(\ref{H-H-Eij}) holds. 
Note that, in the expression of the function $f^2$ in Eq.(\ref{g^2-action}), integration 
is taken over $(\bs{I}_A, \bs{\theta})$. This implies that, for any $(\bm{I}'_A,\bm{\theta}')$, 
the point $(\bm{I}_A,\bm{\theta})$ near $(\bm{I}'_A,\bm{\theta}')$, which makes Eq.(\ref{H-H-Eij}) hold, 
 can always be reached in the integration of Eq.(\ref{g^2-action}).
Such $(\bm{I}_A,\bm{I}'_B,\bm{\theta})$ follows $(\bm{I}', \bm{\theta}')$ running over the whole energy surface of $H=E_i\approx E_j$,
 therefore, the integration  may contain a factor being proportional to $S(E_j)$.
As a consequence, the function $f^2$ may be reversely proportional to $S(E_i) \sim \rho_{\rm dos}$,
as expected in the ETH ansatz.

\acknowledgments

 The authors are grateful to Jiaozi Wang for valuable discussions.
 This work was partially supported by the Natural Science Foundation of China under Grant
 Nos.~12175222, 11535011, and 11775210.

\appendix

\section{Proof of Eq.(\ref{P_Weyl_Modify})}\label{App_P_Weyl_Modify}

 In this appendix, we give details of a proof of Eq.(\ref{P_Weyl_Modify}). 
 The basic strategy is as follows. 
 By making use of the equality of
\begin{equation}
  \frac{1}{2^{n}}\sum_{k=0}^{n}\binom{n}{k}=1,
\end{equation}
 the product is written in the following form, 
\begin{equation}\label{Pmn-1}
  P_{m,n}^{(s)}(\hat{p},\hat{q})=\frac{1}{2^{n}}\sum_{k=0}^{n}\binom{n}{k}P_{m,n}^{(s)}(\hat{p},\hat{q}).
\end{equation}
 For each term on the rhs of Eq.(\ref{Pmn-1}), which corresponds to one fixed value of $k$, 
 by appropriately  permuting $\hat p$ and $\hat q$ in $P_{m,n}^{(s)}(\hat{p},\hat{q})$, 
 one may write it as a sum of $\hat{q}^{n-k}\hat{p}^m\hat{q}^k$ and some other products. 
 According to the definition of Weyl-ordered operator, the sum of 
 $\binom{n}{k}\hat{q}^{n-k}\hat{p}^m\hat{q}^k$ gives rise to a Weyl-ordered operator,
 as the first term on the rhs of Eq.(\ref{P_Weyl_Modify}).

 To be specific, consider a case, in which a product $P_{m,n}^{(s)}(\hat{p},\hat{q})$
 contains a term of $\hat{q}\hat{p}$
 and the above discussed strategy requires an exchange of the positions of 
 $\hat{q}$ and $\hat{p}$ in this term, i.e., $\hat{q}\hat{p} \to \hat{p}\hat{q}$.
 Generically, this product is written as
\begin{equation}
  P_{m,n}^{(s)}(\hat{p},\hat{q})=P_{l_p,l_q}^{(s_L)}(\hat{p},\hat{q}) \, \hat{q}\hat{p}
  \, P_{m-l_p-1,n-l_q-1}^{(s_R)}(\hat{p},\hat{q}),
\end{equation}
 where $l_p$ and $l_q$ are the numbers of $\hat{p}$ and $\hat{q}$, 
 which lie on the left hand side of the term $\hat{q}\hat{p}$.
 The exchange of positions of $\hat q$ and $\hat p$ is achieved by the following relation, 
\begin{equation}\label{Pmn-PLPR-dP}
  \begin{aligned}
    P_{m,n}^{(s)}(\hat{p},\hat{q})
    =&P_{l_p,l_q}^{(s_L)}(\hat{p},\hat{q})\hat{p}\hat{q}P_{m-l_p-1,n-l_q-1}^{(s_R)}(\hat{p},\hat{q})\\
    &+i\hbar P_{l_p,l_q}^{(s_L)}(\hat{p},\hat{q})P_{m-l_p-1,n-l_q-1}^{(s_R)}(\hat{p},\hat{q}).
  \end{aligned}
\end{equation}
 In the opposite case, in which the position exchange required is $\hat{p}\hat{q} \to \hat{q}\hat{p}$, 
 one gets a similar result, but with a minus sign for the second term in an equality
 like Eq.(\ref{Pmn-PLPR-dP}). 

 In both cases, one sees that Eq.(\ref{Pmn-PLPR-dP}) is written as follows, 
\begin{equation}\label{Once_pq_Commutate}
  P_{m,n}^{(s)}(\hat{p},\hat{q})=P_{m,n}^{(s_1)}(\hat{p},\hat{q})+i\hbar d^{(s'_1)} 
  P_{m-1,n-1}^{(s'_1)}(\hat{p},\hat{q}),
\end{equation}
 where $s_1$ and $s_1'$ are labels introduced appropriately and
\begin{equation}\label{}
  d^{(s'_1)} = \left\{
    \begin{array}{ll}
      1, & \hbox{for $\hat{q}\hat{p} \to \hat{p}\hat{q}$} \\
      -1, & \hbox{for $\hat{p}\hat{q} \to \hat{q}\hat{p}$.}
    \end{array}
  \right.
\end{equation}
 The product $P_{m,n}^{(s_1)}(\hat{p},\hat{q})$ may be treated in a similar way, 
\begin{equation}
  P_{m,n}^{(s_1)}(\hat{p},\hat{q})=P_{m,n}^{(s_2)}(\hat{p},\hat{q})+i\hbar d^{(s'_2)} P_{m-1,n-1}^{(s'_2)}(\hat{p},\hat{q}).
\end{equation}
Substitute this result into Eq.(\ref{Once_pq_Commutate}), one gets that
\begin{equation}
  \begin{aligned}
    P_{m,n}^{(s)}(\hat{p},\hat{q})=P_{m,n}^{(s_2)}(\hat{p},\hat{q})
    +&i\hbar d^{(s'_1)} P_{m-1,n-1}^{(s'_1)}(\hat{p},\hat{q})\\
    +&i\hbar d^{(s'_2)} P_{m-1,n-1}^{(s'_2)}(\hat{p},\hat{q}).
  \end{aligned}
\end{equation}

 Repeating the above procedure,  one gets that
\begin{equation}\label{P2kthTerm}
  P_{m,n}^{(s)}(\hat{p},\hat{q})=\hat{q}^{n-k}\hat{p}^m\hat{q}^k
  +i\hbar\sum_{ \{s' \}} d^{(s')} P_{m-1,n-1}^{(s')}(\hat{p},\hat{q}),
\end{equation}
 where $\{s' \}$ indicates the set of those $s'$ that appear in the above procedure. 
 Note that labels in the set $\{s' \}$ are usually $k$-dependent, which is not indicated
 explicitly for brevity.

 Substituting Eq.(\ref{P2kthTerm}) into the rhs of Eq.(\ref{Pmn-1}) 
 with respect to each value of $k$, one gets that
\begin{equation}\label{P2Weyl_1stCommutate}
  \begin{aligned}
    &P_{m,n}^{(s)}(\hat{p},\hat{q})
    =\frac{1}{2^{n}}\sum_{k=0}^{n}\binom{n}{k}\hat{q}^{n-k}\hat{p}^m\hat{q}^k\\
    &\qquad\qquad +\frac{i\hbar}{2^{n}}\sum_{k=0}^{n}\binom{n}{k}\sum_{\{s' \}}
    d^{(s')} P_{m-1,n-1}^{(s')}(\hat{p},\hat{q})\\
    &=P_{m,n}^{\rm Weyl}(\hat{p},\hat{q})
    +\frac{i\hbar}{2^{n}}\sum_{k=0}^{n}\binom{n}{k}\sum_{\{s' \}}d^{(s')} P_{m-1,n-1}^{(s')}(\hat{p},\hat{q}).
  \end{aligned}
\end{equation}

 Below, we deal with the second part on the rhs of Eq.(\ref{P2Weyl_1stCommutate}). 
 In fact, each term $P_{m-1,n-1}^{(s')}(\hat{p},\hat{q})$ may be treated in a way similar 
 to that discussed above for $P_{m,n}^{(s)}(\hat{p},\hat{q})$. 
 This implies that
\begin{equation}
  \begin{aligned}
    P_{m-1,n-1}^{(s')}&(\hat{p},\hat{q})=P_{m-1,n-1}^{\rm Weyl}(\hat{p},\hat{q})\\
    &+\frac{i\hbar}{2^{n-1}}\sum_{k_2=0}^{n-1}\binom{n-1}{k_2}
    \sum_{\{s'' \}}d^{(s'')} P_{m-2,n-2}^{(s'')}(\hat{p},\hat{q}).
  \end{aligned}
\end{equation}

 Repeating the above procedure, one gets the following expression of $P_{m,n}^{(s)}(\hat{p},\hat{q})$,
\begin{equation}\label{P2Weyl_lastCommutate}
  \begin{aligned}
    &P_{m,n}^{(s)}(\hat{p},\hat{q})
    =P_{m,n}^{\rm Weyl}(\hat{p},\hat{q})\\
    &\qquad +\sum_{\lambda=1}^{M-1} (i\hbar)^\lambda D_{m-\lambda,n-\lambda}^{(s)} P^{\rm Weyl}_{m-\lambda,n-\lambda}(\hat{p},\hat{q})\\
    &+\frac{(i\hbar)^M}{2^{(2n+1+M)M/2}}\prod_{\rho=1}^M\left[\sum_{k_\rho=0}^{n+1-\rho}\binom{n+1-\rho}{k_\rho}\right.\\
    &\qquad \left.\times \sum_{\{s'_\rho \} }d^{(s'_\rho)}\right]P_{m-M,n-M}^{(s'_M)}(\hat{p},\hat{q}),
  \end{aligned}
\end{equation}
where $M=\min(m,n)$  and
\begin{equation}\label{D_Coefficient}
  \begin{aligned}
    D_{m-\lambda,n-\lambda}^{(s)}&=
    \frac{1}{2^{(2n+1+\lambda)\lambda/2}}\\
    &\times\prod_{\rho=1}^\lambda\left[\sum_{k_\rho=0}^{n+1-\rho}\right.
    \left.\binom{n+1-\rho}{k_\rho}\sum_{\{ s'_\rho \}}d^{(s'_\rho)}\right].
  \end{aligned}
\end{equation}
 Here, dependence of $d^{(s'_\rho)}$ on $k_\rho$ and $s'_{\rho-1}$ is not written explicitly for brevity.

For $P_{m-M,n-M}^{(s'_M)}(\hat{p},\hat{q})$, by the definition of $M$, we know 
$m-M=0$ or $n-M=0$, which means $P_{m-M,n-M}^{(s'_M)}(\hat{p},\hat{q})=\hat{q}^{n-M}$ 
or $\hat{p}^{m-M}$. The sequence is fixed, so $s'_M$ has only one permitted 
value. By the way, in both case, according to the definition Eq.(\ref{P_WeylOrder}),
$P_{m-M,n-M}^{(s'_M)}(\hat{p},\hat{q})=P_{m-M,n-M}^{\rm Weyl}(\hat{p},\hat{q})$.
Therefore, the $P_{m-M,n-M}^{(s'_M)}(\hat{p},\hat{q})$ can be moved outside 
the summation. Finally we have
\begin{equation}\label{P_Weyl_Modify_App}
  \begin{aligned}
    P_{m,n}^{(s)}&(\hat{p},\hat{q})
    =P_{m,n}^{\rm Weyl}(\hat{p},\hat{q})\\
    &+\sum_{\lambda=1}^{M} (i\hbar)^\lambda D_{m-\lambda,n-\lambda}^{(s)} P^{\rm Weyl}_{m-\lambda,n-\lambda}(\hat{p},\hat{q}),
  \end{aligned}
\end{equation}
where $D_{m-\lambda,n-\lambda}^{(s)}$ are also given in Eq.(\ref{D_Coefficient}).

It is Eq.(\ref{P_Weyl_Modify}). 

\bibliography{ETH-sc.bbl}

\end{document}